\documentclass[useAMS,usenatbib]{mn2e}

\usepackage{times,graphicx,amssymb,natbib}

\title[Cloud Angular Momentum and Disc Fragmentation]{A Lower Angular Momentum Limit for Self-Gravitating Protostellar Disc Fragmentation}
\author[Duncan Forgan and Ken Rice]{Duncan Forgan $^{1}$\thanks{E-mail:dhf@roe.ac.uk} and Ken Rice$^{1}$ \\
$^{1}$Scottish Universities Physics Alliance (SUPA), Institute for Astronomy, University of Edinburgh, Blackford Hill, Edinburgh, EH9 3HJ, Scotland, UK}
\begin{document}

\date{Accepted}

\pagerange{\pageref{firstpage}--\pageref{lastpage}} \pubyear{}

\maketitle

\label{firstpage}

\begin{abstract}

\noindent We attempt to verify recent claims (made using
semi-analytic models) that for the collapse of spherical homogeneous
molecular clouds, fragmentation of the self-gravitating disc that
subsequently forms can be predicted using the cloud's initial angular
momentum alone.  In
effect, this condition is equivalent to requiring the resulting disc
be sufficiently extended in order to fragment, in line with studies of
isolated discs.  We use smoothed particle hydrodynamics with hybrid
radiative transfer to investigate this claim, confirming that in
general, homogeneous spherical molecular clouds will produce
fragmenting self-gravitating discs if the ratio of rotational kinetic energy
to gravitational potential energy is greater than $\approx 5\times 10^{-3}$,
where this result is relatively insensitive to the initial thermal
energy. This condition begins to fail at higher cloud masses,
suggesting that sufficient mass at large radii governs
fragmentation. While these results are based on highly idealised
initial conditions, and may not hold if the disc's accretion from the
surrounding envelope is sufficiently asymmetric, or if the density structure is perturbed, they provide a
sensible lower limit for the minimum angular momentum required to
fragment a disc in the absence of significant external turbulence.

\end{abstract}


\section{Introduction}

\noindent Cold, dense molecular cloud cores are thought to be the sites of low-mass star formation \citep{Terebey1984}.  Typically, these cores will contain an excess of angular momentum (compared with the rotational angular momentum of low-mass stars, cf \citealt{Caselli2002}).  If these cores are sufficiently dense, they will collapse under their own gravity to form a protostar plus protostellar disc, with a substantial envelope surrounding both.  In this pre-main sequence phase, we can expect that most of the star's eventual mass will be processed by the protostellar disc (which in turn accretes this mass from a surrounding envelope).  

The star's accretion is inextricably linked with the process of outward angular momentum transport - understanding the mechanisms by which angular momentum transport functions in these early phases is therefore essential to theories of star formation.  While turbulence induced by the magneto-rotational instability (MRI) defines angular momentum transport at late times \citep{Blaes1994,Balbus1999}, the disc is unlikely to be even weakly ionised at early times.  We must therefore seek another means by which to generate angular momentum transport.  

Simulations suggest that the disc is likely to be self-gravitating at these early times, as the disc mass will be large relative to the star mass \citep{Bate2010, Machida2010}.  If the disc becomes gravitationally unstable, this may provide the dominant transport mechanism through the growth of self-regulated gravito-turbulence \citep{Boss1984,Lin1987,Laughlin1994}.  A self-gravitating disc will become subject to gravitational instability (GI) if the Toomre parameter, $Q$ \citep{Toomre_1964}:

\begin{equation} Q = \frac{c_s \kappa}{\pi G \Sigma} \sim 1, \end{equation}

\noindent where $c_s$ is the sound speed, $\Sigma$ is the surface density, and $\kappa$ is the epicyclic frequency (in the case of a Keplerian disc, $\kappa$ is equal to the angular frequency $\Omega$).  If $Q<1$, axisymmetric perturbations will grow within the disc.  Numerical simulations have established that discs with $Q<1.5-1.7$ will be unstable to non-axisymmetric perturbations \citep{durisen_review}.  The onset of GI is accompanied by the growth of quasi-steady spiral structures, which drive shocks with Mach numbers of order unity \citep{Cossins2008}.  

The growth of GI in a self-gravitating disc will generally lead to a
state commonly referred to as marginal instability
\citep{Paczynski1978}, where the heating due to GI is balanced by
radiative cooling.  The combination of these two processes produces a
quasi-equilibrium state where the instability's amplitudes are
modulated, providing a self-regulated outward angular momentum
transport process \citep{Gammie,Mejia_1,Lodato_and_Rice_04,Mejia_4}.  The growth and decay of these structures
provides the so-called gravito-turbulence, allowing the evolution of
the disc to be described in a pseudo-viscous fashion
(e.g. \citealt{Balbus1999,Lodato_and_Rice_04,Rice_and_Armitage_09,Clarke_09}), where the transport
can be characterised using the dimensionless turbulent viscosity
parameter $\alpha$ \citep{Shakura_Sunyaev_73}.  While this local $\alpha$-parametrisation breaks down in certain limits \citep{Forgan2011}, it remains a useful tool to compare to other transport mechanisms such as the MRI. 

Self-gravitating disc evolution is also an important component of planet formation theory, not merely through its effects on the growth and accretion of planetary cores \citep{Rice2004,Clarke2009}.  If the disc is self-gravitating, its dynamical evolution can rapidly form giant planets and brown dwarfs by disc fragmentation \citep{Kuiper1951,Boss_science,Stam_frag}.  While the $Q$ criterion is a necessary condition for disc fragmentation, it is insufficient.  A second criterion is required, relating to the local cooling time:

\begin{equation} t_{\rm cool} = \frac{u}{du/dt}, \end{equation}

\noindent where $u$ is the specific internal energy.  It is often more suitable to discuss cooling in terms of a dimensionless parameter, $\beta_{c}$:

\begin{equation} \beta_{c} = t_{\rm cool} \Omega. \end{equation}

\noindent  \citet{Gammie} showed using shearing-sheet simulations that if $\beta_c = \mathrm{const}.$, fragmentation can only occur if $\beta_c < \beta_{\rm crit}$, where $\beta_{\rm crit}$ is approximately 3, although it was later shown that $\beta_{\rm crit}$ depends on the local equation of state, in particular the value of the ratio of specific heats, $\gamma$ \citep{Rice_et_al_05}. The correct values of $\beta_{\rm crit}(\gamma)$ have been challenged by recent work, and may be slightly higher \citep{Meru2011,Paardekooper2011}.  In the case of smoothed particle hydrodynamics (SPH) simulations as are carried out in this work, this appears to be due to resolution effects pushing the convergence limit to higher particle number than was previously anticipated \citep{Lodato2011}, as well as possible issues regarding the correct implementation of cooling prescriptions (Rice, Forgan and Armitage, submitted).

Alternatively, the second fragmentation criterion can be couched in
terms of the local stresses in the disc.  In the Shakura-Sunyaev
formalism, the effective turbulent viscosity $\nu = \alpha c_s H$,
where $H$ is the disc scale height.  Stresses have two principal
origins in self-gravitating discs\footnote{A third source of disc
  stress is magnetic fields, but as the self-gravitating phase is typically very weakly ionised, the magnitude of these stresses is small, and hence we do not consider them here} - Reynolds stresses (induced by correlated velocity perturbations) and gravitational stresses (induced by correlated perturbations in the gravitational potential).  Assuming that the disc is in thermal equilibrium, the value of $\alpha$ can be determined by balancing the pseudo-viscous heating and the radiative cooling:

\begin{equation} \alpha_{\rm cool} = \left(\frac{d \ln \Omega}{d \ln r}\right)^{-2} \frac{1}{\gamma(\gamma-1)\beta_{c}}. \label{eq:alpha_tcool}\end{equation}

\noindent \citet{Rice_et_al_05} show that a single critical value of
$\alpha=\alpha_{\rm crit} \approx 0.06$ can be used as the second fragmentation criterion, for
any value of $\gamma$.  If the disc is irradiated, it has been shown
that $\alpha$, does not uniquely determines the likelihood of
fragmentation, and that short cooling times can induce fragmentation
even in low-stress environments \citep{Rice2011}.  Studies have however
shown that this $\alpha_{\rm crit}$ value may not apply if $\beta_c$ can vary \citep{Clarke2007}, or if the disc's accretion from the envelope is substantial \citep{Kratter2008,Harsono2011}.  In these circumstances, more sophisticated criteria are required (see e.g. \citealt{Forgan2011a}).  In spite of these circumstances, it has been established by many authors that the requisite conditions for fragmentation only occur in the outer regions of discs \citep{Rafikov_05, Matzner_Levin_05, Whit_Stam_06,Mejia_3,Stamatellos2008, intro_hybrid, Clarke_09, Rice2010}.  These results hold in semi-analytic calculations, grid-based simulations and particle-based simulations.  This does not limit the usefulness of disc fragmentation as an explanation for observable exoplanets - indeed, it explains the properties of systems such as HR 8799 when core accretion theories struggle to do so \citep{Kratter2010,Nero2009}.  \citet{Meru2011a} presented SPH simulations which displayed inner disc fragmentation for steep surface density profiles, but it has been shown that these can be explained by failure to address resolution criteria relating to the artificial viscosity \citep{Lodato2011} - indeed, a failure that all previous SPH simulations of this type have shared.

As has been stated, the self-gravitating phase is expected to occur at
early times, when the disc is still embedded in the envelope.
Therefore, to fully understand the fragmentation of self-gravitating
discs, it is necessary to include the envelope's contribution, both in
its ability to supply the disc with matter \citep{Kratter2008}, and in
the resulting stresses it can induce at the disc's surface and beyond
\citep{Harsono2011}.  These stresses result in extensive non-local
angular momentum transport, a phenomenon which will tend to invalidate semi-analytic models \citep{Forgan2011}.  

A straightforward means by which the envelope can be included is to begin with a rotating molecular cloud as initial conditions.  Frequently, the cloud parameters $\phi$ and $\lambda$ are used\footnote{In the literature, these parameters are known as $\alpha$ and $\beta$ respectively.  We relabel them to avoid confusion with the subsequent disc parameters.} where:

\begin{equation} \phi = \frac{E_{\rm therm}}{E_{\rm grav}} \end{equation}
\begin{equation} \lambda = \frac{E_{\rm rot}}{E_{\rm grav}},\end{equation}

\noindent where $E_{\rm therm}$, $E_{\rm rot}$ and $E_{\rm grav}$ are
the cloud's total thermal, rotational and gravitational energies
respectively.  For an initially isothermal cloud in rigid body rotation, these parameters essentially define the cloud's initial temperature and angular velocity:

\begin{equation} \phi = \frac{5kT_0R_{\rm cl}}{2\mu m_H G M_{\rm cl}} \end{equation}
\begin{equation} \lambda = \frac{\Omega_0^2 R_{\rm cl}^3}{3 G M_{\rm cl}} ,\end{equation}

\noindent where \(R_{\rm cl}\) is the cloud's initial radius, \(k\) is Boltzmann's constant, \(\mu\) is the mean molecular weight of the gas and \(m_H\) is the mass of the hydrogen atom.  \citet{Tohline1981} considers the collapse of adiabatic spheroids (\(P = K\rho^{\Gamma}\)), and defines a series of criteria for these clouds to become unstable to nonaxisymmetric perturbations, dependent on the polytropic index \(\Gamma\) of the cloud:

\begin{equation} \phi \leq \left\{
\begin{array}{l l }
0.109\,\lambda ^{-1} & \quad \mbox{\(\Gamma = 1\)} \\
0.201 & \quad \mbox{\(\Gamma = 4/3\)} \\
0.228\,\lambda^{0.2} & \quad \mbox{\(\Gamma = 7/5\)} \\
0.372\,\lambda &\quad \mbox{\(\Gamma = 5/3.\)} \\
\end{array} \right. \end{equation}

\noindent In reality, a collapsing molecular cloud will have a more
complex thermal history that requires a more detailed understanding of
the equation of state as a function of the local density and
temperature. \citet{Rice2010}, using a model initially developed by
\citet{Lin1990}, carried out semi-analytic calculations of
uniform, spherical molecular cloud collapse, with a more complex
equation of state, using the pseudo-viscous approximation to evolve
the resulting disc, which continues to accrete from its envelope. They
show that the resulting self-gravitating discs are subject to
fragmentation (i.e. they contain regions where $\alpha_{\rm cool}>
0.06$) when $\lambda> 1\times10^{-3}$, and never fragment when
$\lambda \leq 1\times10^{-3}$, \emph{regardless of the initial thermal
  energy of the cloud}.  

It is not clear whether these results can be
expected to hold, given that non-local angular momentum transport
(which must play a role, especially at early times) is not accounted
for, as well as potentially asymmetric accretion flows.  In addition,
these initial conditions are extremely simple, and not likely to be
reproduced in reality - a more appropriate choice would be
supercritical Bonnor-Ebert spheres (e.g. \citealt{Walch2009}), as well
as the addition of turbulence \citep{Walch2010}.  Finally, as has been
previously stated, using $\alpha_{\rm crit} = 0.06$ is probably not
appropriate for accreting discs.

We aim to perform a set of numerical experiments using smoothed
particle hydrodynamics with hybrid radiative transfer
\citep{intro_hybrid} to appraise the validity of the angular momentum
condition set forth by \citet{Rice2010}.  Unlike \citet{Walch2009}, we
do not hold $\phi$ constant, and therefore we can also detect whether
any dependence on thermal energy exists.  Also, we can model non-local
angular momentum transport, and can measure fragmentation directly
rather than relying on uncertain criteria, allowing us to test more
reliably whether an angular momentum criterion truly exists.

\section{Method }\label{sec:Method}

\subsection{SPH and the Hybrid Radiative Transfer Approximation}

\noindent Smoothed Particle Hydrodynamics (SPH) \citep{Lucy,Gingold_Monaghan,Monaghan_92} is a Lagrangian formalism that represents a fluid by a distribution of particles.  Each particle is assigned a mass, position, internal energy and velocity.  State variables such as density and pressure are then calculated by interpolation (see reviews by \citealt{Monaghan_92,Monaghan_05}).  In the simulations presented here, the gas is modelled using 500,000 SPH particles while the star is represented by a point mass particle onto which gas particles can accrete, if they are sufficiently close (in our case, within $1$ au) and are bound \citep{Bate_code}.  An SPH particle is converted to a pointmass if the local density has exceeded $10^{-12}\, \rm{g\, cm}^{-3}$, and

\begin{enumerate}
\item There exists one Jeans mass within the particle's neighbour sphere
\item The neighbour sphere is bound (i.e. the total energy inside the sphere is negative)
\item The local gravitational potential energy exceeds both the thermal and rotational kinetic energy
\end{enumerate}

The SPH code used in this work is based on the SPH code developed by \citet{Bate_code} which uses individual particle timesteps, and individually variable smoothing lengths, $h_{\rm i}$, such that the number of nearest neighbours for each particle is \(50 \pm 20\).  The code uses a hybrid method of approximate radiative transfer \citep{intro_hybrid}, which is built on two pre-existing radiative algorithms: the polytropic cooling approximation devised by \citet{Stam_2007}, and flux-limited diffusion (e.g., \citealt{WB_1,Mayer2007}, see \citealt{intro_hybrid} for details).  This union allows the effects of both global cooling and radiative transport to be modelled, without imposing extra boundary conditions. 

The opacity and temperature of the gas is calculated using a non-trivial equation of state.  This accounts for the effects of H$_{\rm 2}$ dissociation, H$^{\rm 0}$ ionisation, He$^{\rm 0}$ and He$^{\rm +}$ ionisation, ice evaporation, dust sublimation, molecular absorption, bound-free and free-free transitions and electron scattering \citep{Bell_and_Lin,Boley_hydrogen,Stam_2007}.  Heating of the disc is also achieved by \(P\,dV\) work and shock heating.

\subsection{Initial Cloud Conditions}

\noindent The molecular clouds were initialised with $500,000$ SPH particles.  While this may seem under-resolved in the light of the results from \citet{Meru2011}, it is important to note that they use a fundamentally different cooling method where particle cooling rates are not affected by their neighbours.  The hybrid method, through its flux limited diffusion component, does account for the neighbour sphere.  A more appropriate resolution criterion (\citealt{Lodato2011}, based on the results of \citealt{Meru2011}) shows that 500,000 particles is sufficient to resolve fragmentation, and is discussed in the following section.
 The particles were distributed in a sphere with uniform density.  The clouds were initialised with varying values of cloud mass $M_{cl}$ and $\lambda$ - the cloud radii $R_{cl}$ were specified in order to maintain the same initial density $\rho_0 = 1.46 \times 10^{-17}\, \mathrm{g \,cm^{-3}}$.  This ensures that all clouds have the same free-fall time

\begin{equation} t_{ff} = \sqrt{\frac{3\pi}{32 G\rho_0}} = 1.74 \times 10^{4} \,\mathrm{yr}.  \end{equation} 

\noindent The initial cloud temperature is fixed at 5 K, equal to the background temperature (beyond which the cloud cannot cool).  Note by fixing the parameters in this fashion, we allow $\phi$ to vary.  We choose this setup as it is similar to that of the semi-analytic calculations of \citet{Rice2010}, who also allow $\phi$ to vary in this fashion.  Their results appear to be independent of $\phi$, and we wish to assess if this is true in our case.  We do not apply a density perturbation to the clouds, as is often done in simulations of this type (e.g. \citealt{Boss1986}).  The amplitude of the $m=2$ mode initially is zero in our simulations - the subsequent growth of $m=2$ spiral modes is purely a result of disc torques forming from gravitational instabilities.

\begin{table}
\centering
\begin{minipage}{140mm}
  \caption{Summary of the cloud parameters investigated in this work.\label{tab:params}}
  \begin{tabular}{c || ccccc}
  \hline
  \hline
   Simulation & $M_{\rm cl}$ ($M_{\rm \odot}$) & $R_{\rm cl}$ (au) & $\lambda$ &$\phi$   \\  
 \hline  
  1 & 1.0 & 2133.33 & 4 $\times 10^{-3}$ & 0.2 \\
  2 & 1.0 & 2133.33 & 5 $\times 10^{-3}$ & 0.2  \\
  3 & 1.0 & 2133.33 & 6 $\times 10^{-3}$ & 0.2  \\
  4 & 1.0 & 2133.33 & 7 $\times 10^{-3}$ & 0.2 \\
  5 & 1.0 & 2133.33 & 8 $\times 10^{-3}$ & 0.2 \\
  6 & 1.0 & 2133.33 & 9 $\times 10^{-3}$ & 0.2 \\
  \hline
  7 & 0.5 & 1693.23 & 4 $\times 10^{-3}$ & 0.4 \\
  8 & 0.5 & 1693.23 & 5 $\times 10^{-3}$ & 0.4 \\
  9 & 0.5 & 1693.23 & 6 $\times 10^{-3}$ & 0.4 \\
  10 & 0.5 & 1693.23 & 7 $\times 10^{-3}$ & 0.4 \\
  11 & 0.5 & 1693.23 & 8 $\times 10^{-3}$ & 0.4 \\
  12 & 0.5 & 1693.23 & 9 $\times 10^{-3}$ & 0.4 \\
  \hline
  13 & 2.0 & 2687.83 & 4 $\times 10^{-3}$ & 0.1 \\
  14 & 2.0 & 2687.83 & 5 $\times 10^{-3}$ & 0.1 \\
  15 & 2.0 & 2687.83 & 6 $\times 10^{-3}$ & 0.1  \\
  16 & 2.0 & 2687.83 & 7 $\times 10^{-3}$ & 0.1 \\
  17 & 2.0 & 2687.83 & 8 $\times 10^{-3}$ & 0.1 \\
  18 & 2.0 & 2687.83 & 9 $\times 10^{-3}$ & 0.1 \\
 \hline
  \hline
\end{tabular}
\end{minipage}
\end{table}

\subsection{Resolution Conditions}

\noindent  To ensure that potential fragmentation is resolved, the minimum resolvable Jeans mass - one neighbour group of SPH particles \citep{Burkert_Jeans}, around 50 in the case of the code used - must be sufficiently small:

\begin{equation} M_{\rm min} = 2 N_{\rm neigh}m_{\rm i} = 2 M_{\rm tot} \frac{N_{\rm neigh}}{N_{\rm tot}}. \end{equation}

\noindent The minimum resolvable Jeans mass ranges between $33 M_{\rm \oplus}$ for $M_{\rm cl} = 0.5 M_{\odot}$ to $120 M_{\rm \oplus}$ for $M_{\rm cl} = 2 M_{\odot}$.  As it is expected that fragment masses will be typically several orders of magnitude higher than these values \citep{Kratter2010,Forgan2011a}, this establishes that the simulations would comfortably resolve disc fragmentation if it were to occur.

Perhaps more important are the resolution issues raised by artificial viscosity \citep{Lodato2011}.  While required by the SPH code, this artificial viscosity must be 
quantified so that we know where in the disc the artificial viscosity is likely to be lower than the effective viscosity generated 
by the gravitational instabilities.  If the stress induced by artificial viscosity exceeds that of the stresses induced by the gravitational instability in some region of the disc, then fragmentation is likely to be suppressed \citep{Whitworth1998}.

The linear term for the artificial viscosity can be expressed as \citep{Artymowicz1994,Murray1996,Lodato2010}

\begin{equation} \nu_{\rm art} = \frac{1}{10} \alpha_{\rm SPH} c_{\rm s} h, \label{eq:nu_art}\end{equation}

\noindent where $c_{\rm s}$ is the local sound speed, $h$ is the local SPH smoothing length, and $\alpha_{\rm SPH}$ is the linear viscosity coefficient used 
by the SPH code (taken to be 0.1). We can define an effective $\alpha$ parameter associated with the artificial viscosity \citep{Lodato_and_Rice_04,Forgan2011}

\begin{equation} \nu_{\rm art} = \alpha_{\rm art} c_{\rm s} H, \label{eq:ss_art} \end{equation}

\noindent and hence combining equations (\ref{eq:nu_art}) and (\ref{eq:ss_art}) gives \citep{Artymowicz1994,Murray1996,Lodato2010}

\begin{equation} \alpha_{\rm art} = \frac{1}{10} \alpha_{\rm SPH} \frac{h}{H}. \end{equation}

\noindent This shows that where the vertical structure is not well resolved (i.e., $\frac{h}{H}$ is large), artificial viscosity will dominate.  In the simulations presented here, this is likely to be the case inside $\sim 10$ au, so any data inside this region can not be used.  We therefore only consider results outside this radius, and therefore cannot comment on works which demonstrate fragmentation inside this radius.  In the regions where fragmentation is found, the artificial viscosity is less than 1\% of the total, and therefore satisfies the ``less than 5\%'' criterion given by \citet{Lodato2011} for correctly resolving fragmentation.

Finally, we must recognise that these simulations contain Poisson
noise in the initial conditions.  After approximately one free-fall
time, the star-disc system begins to settle into a marginally unstable
state, effectively erasing the initial conditions.  Fragmentation due
to disc instability typically takes an extra $0.5 t_{ff}$ after disc
formation, corresponding to many outer rotation periods (ORPs), so we do not expect this initial noise to significantly alter our result.  To confirm this, all simulations were repeated, using a different set of randomly generated initial conditions, to confirm that this initial noise does not affect the final result. 

\section{Results}

\subsection{General Trends}

\noindent Each of the clouds goes through a set of generic
evolutionary phases, similar in nature to those described by
\citet{Masunaga_1} for a non-rotating homogeneous cloud.  Initially,
the collapse is isothermal - despite the central density increasing by
several orders of magnitude, the central temperature remains constant
at the background temperature of 5K.  As the central density increases
past $10^{-12} \mathrm{\,g\,cm^{-3}}$, the gas becomes optically
thick, and the central temperature begins to rise.  As the temperature
rises, rotational degrees of freedom in $H_2$ are activated, and the
contraction slows somewhat as the cloud develops significant thermal
pressure support, forming the ``first core'' (cf
\citealt{Larson1969}).    As these simulations contain a sufficient
amount of initial angular momentum, this oblate first core will
subsequently form the protostellar disc \citep{Machida2010}.  This
process begins with the formation of a bar within the core, which
decays into spiral structures (cf \citealt{Bate2010}).  The second
core forms at the centre of this disc structure at approximately 2000
K, when $H_2$ begins to dissociate, removing thermal energy and hence
pressure support. At this stage, the simulation forms a pointmass. The
disc-to-pointmass mass ratio is initially larger than 3, and the star
accretes rapidly.  After this rapid accretion phase, the mass ratio
drops below unity, and the marginally unstable self-gravitating phase begins (except in the simulations that produce multiple discs). 

\begin{figure}
\begin{center}
\includegraphics[scale=0.5]{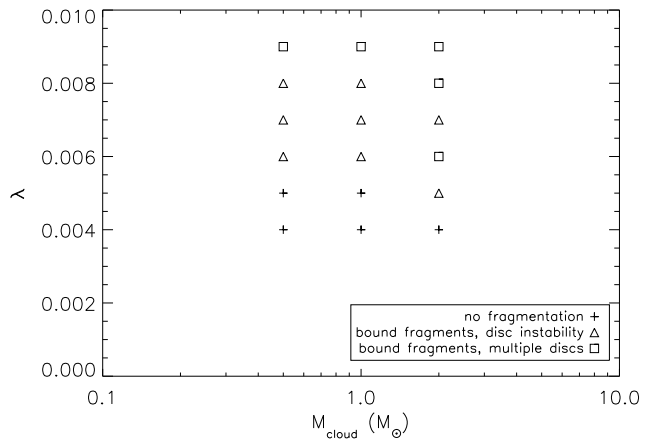}
\caption{The fragmentation boundary in the simulations conducted.  The boundary between disc fragmentation and no fragmentation depends only weakly on initial cloud mass. \label{fig:frag_betavsM}}
\end{center}
\end{figure}

\noindent Figure \ref{fig:frag_betavsM} displays the end state of all the simulations run in this work.  The end states can be categorised into one of three broad classes:

\begin{enumerate}
\item Non-fragmenting, marginally unstable self-gravitating discs, (denoted by '+' in Figure \ref{fig:frag_betavsM}), 
\item Self-gravitating discs which fragment several ORPs after formation (denoted by triangles in Figure \ref{fig:frag_betavsM}),
\item Systems which fragment into multiple systems before a single disc system can form (denoted by squares in Figure \ref{fig:frag_betavsM}).
\end{enumerate}

\noindent It can be seen that the delineation of the three classes shows only a very weak dependence on the initial cloud mass $M_{cl}$.  Note that all clouds have the same initial free-fall time, and the same external background temperature.  The clouds are initially isothermal, and will therefore cool quickly to the background temperature, removing any dependence on initial thermal energy (unlike the adiabatic clouds of for example \citealt{Tohline1981}) The fact that initial angular momentum provides a (relatively) simple criterion 

\begin{equation} \lambda >  5 \times 10^{-3} \end{equation}

\noindent (and its subsequent weak violation at larger cloud masses) suggests that fragmentation requires placing a sufficient quantity of material at large radii at early times, having the resulting effect of reducing the $Q$ parameter without triggering a strong increase in the local cooling time.  This allows sufficient stress to be generated as a result of the gravitational instability, which fragments the disc.  Our results contrast with those from simulations where an azimuthal density perturbation is applied, for example \citet{Boss1986}, who find that binary formation occurs at $\lambda=4\times 10^{-3}$.  This discrepancy seems reasonable again under the interpretation of sufficient mass at large radii at early times promoting fragmentation. 

Discs which do not fragment are qualitatively different in their properties to discs that do.  These differences are generic to all initial cloud masses.  For the sake of brevity, we will only explore the $M_{\rm cl}=1M_{\odot}$ clouds in detail.

\subsection{The $M_{cl} = 1 M_{\odot}$ Simulations} 

\subsubsection{General Properties}

\begin{figure*}
\begin{center}$
\begin{array}{cc}
\includegraphics[scale=0.4]{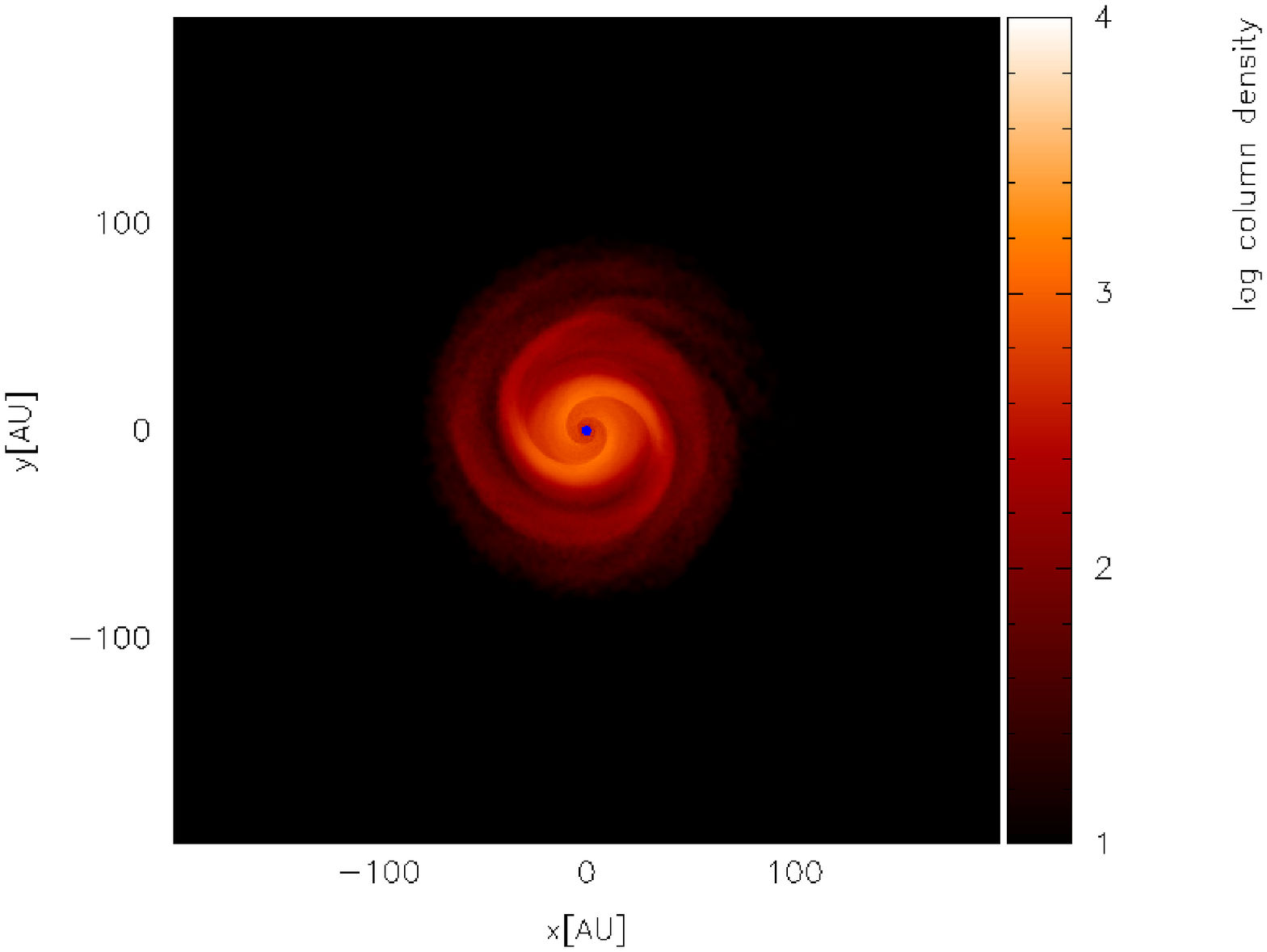} &
\includegraphics[scale = 0.4]{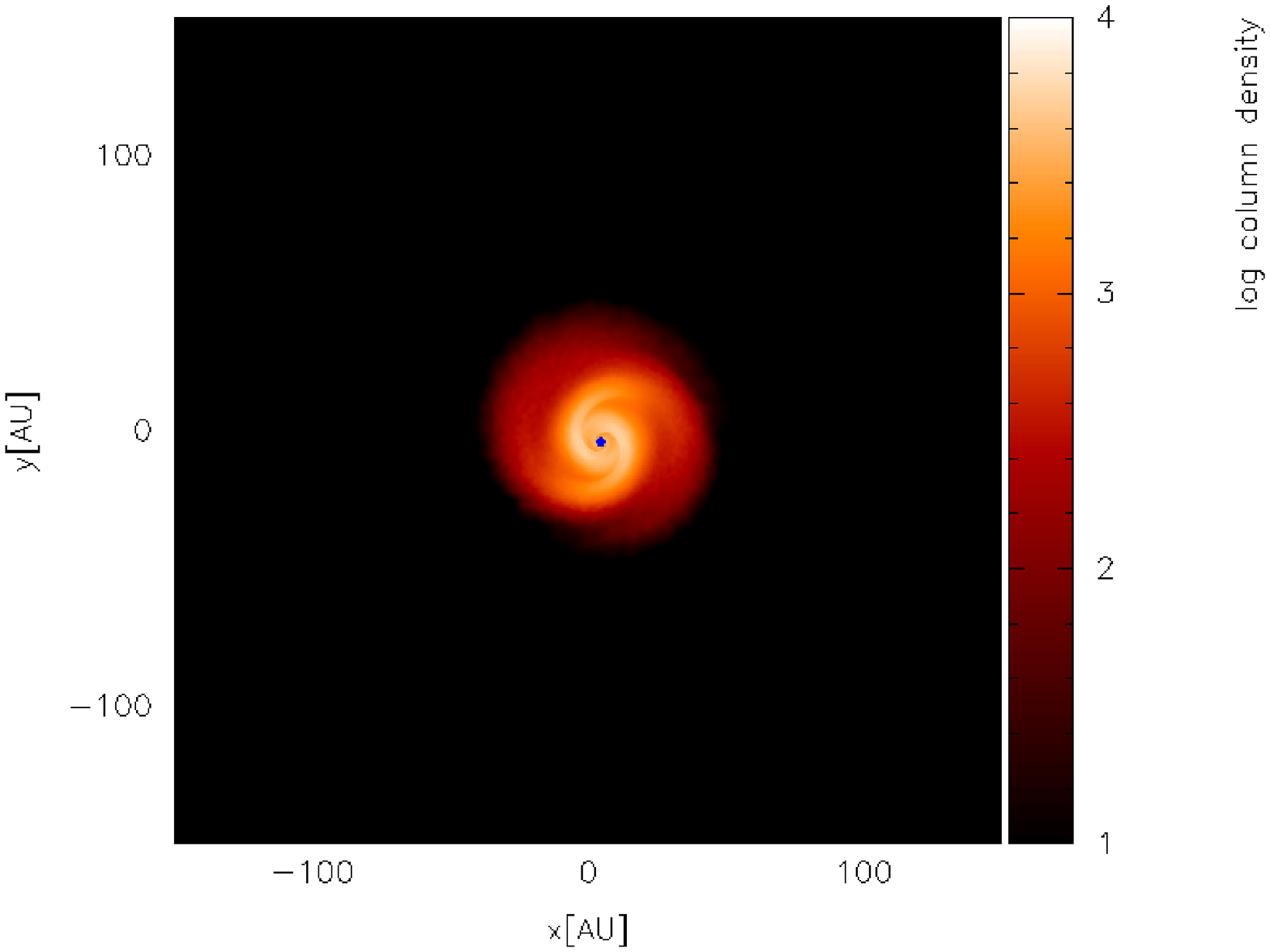} \\
\includegraphics[scale = 0.4]{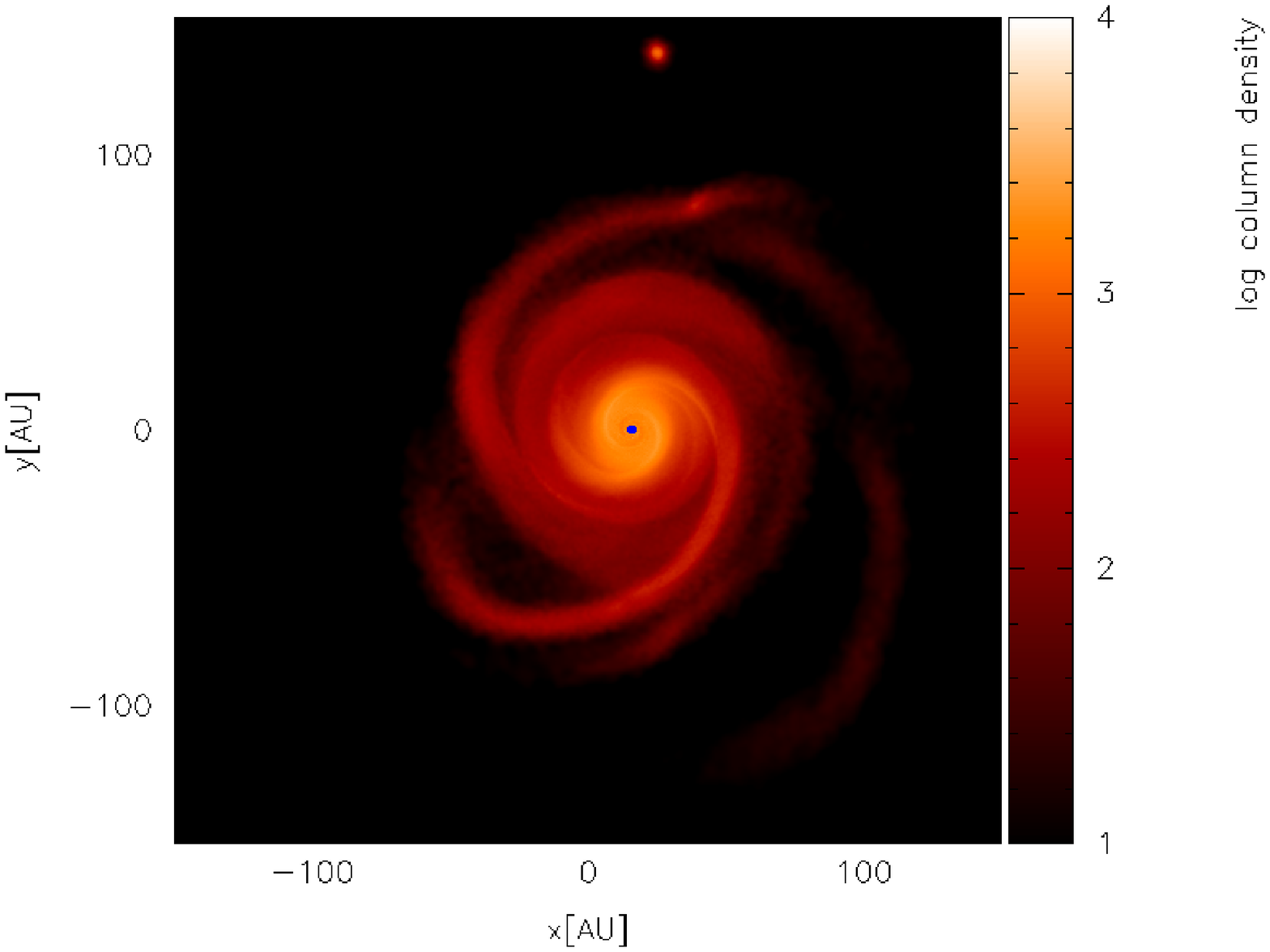} &
\includegraphics[scale = 0.4]{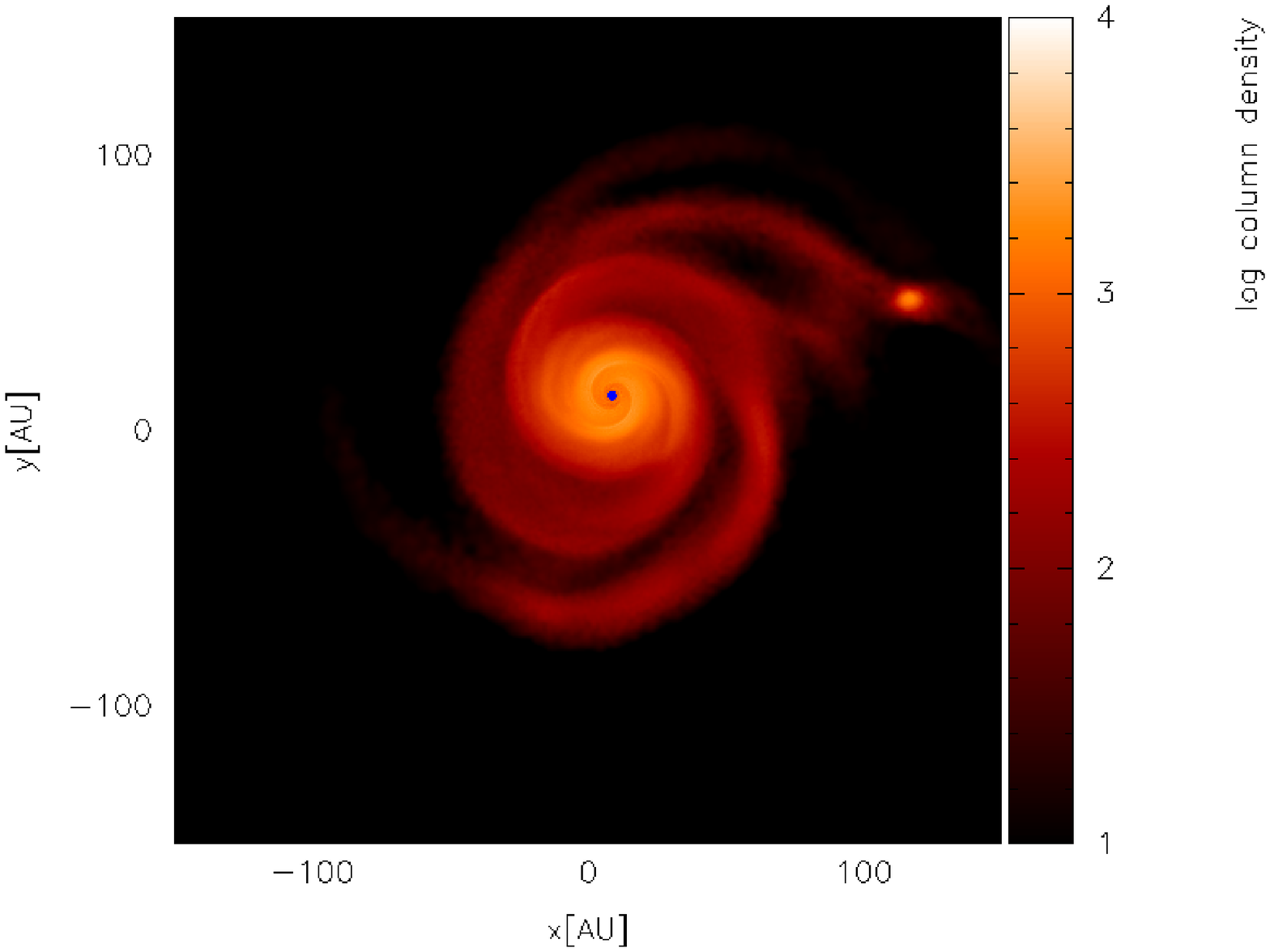} \\
\includegraphics[scale = 0.4]{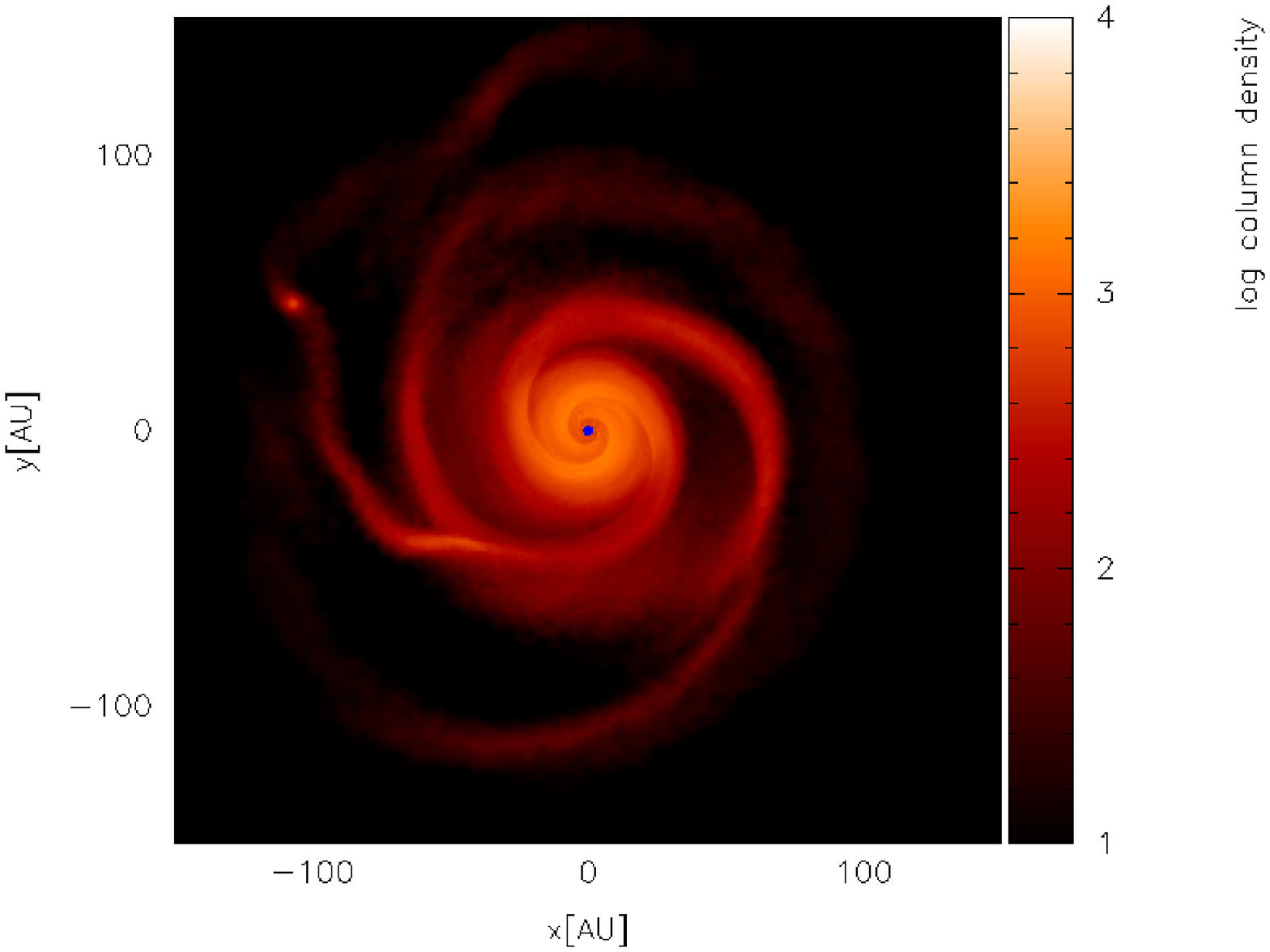} &
\includegraphics[scale=0.4]{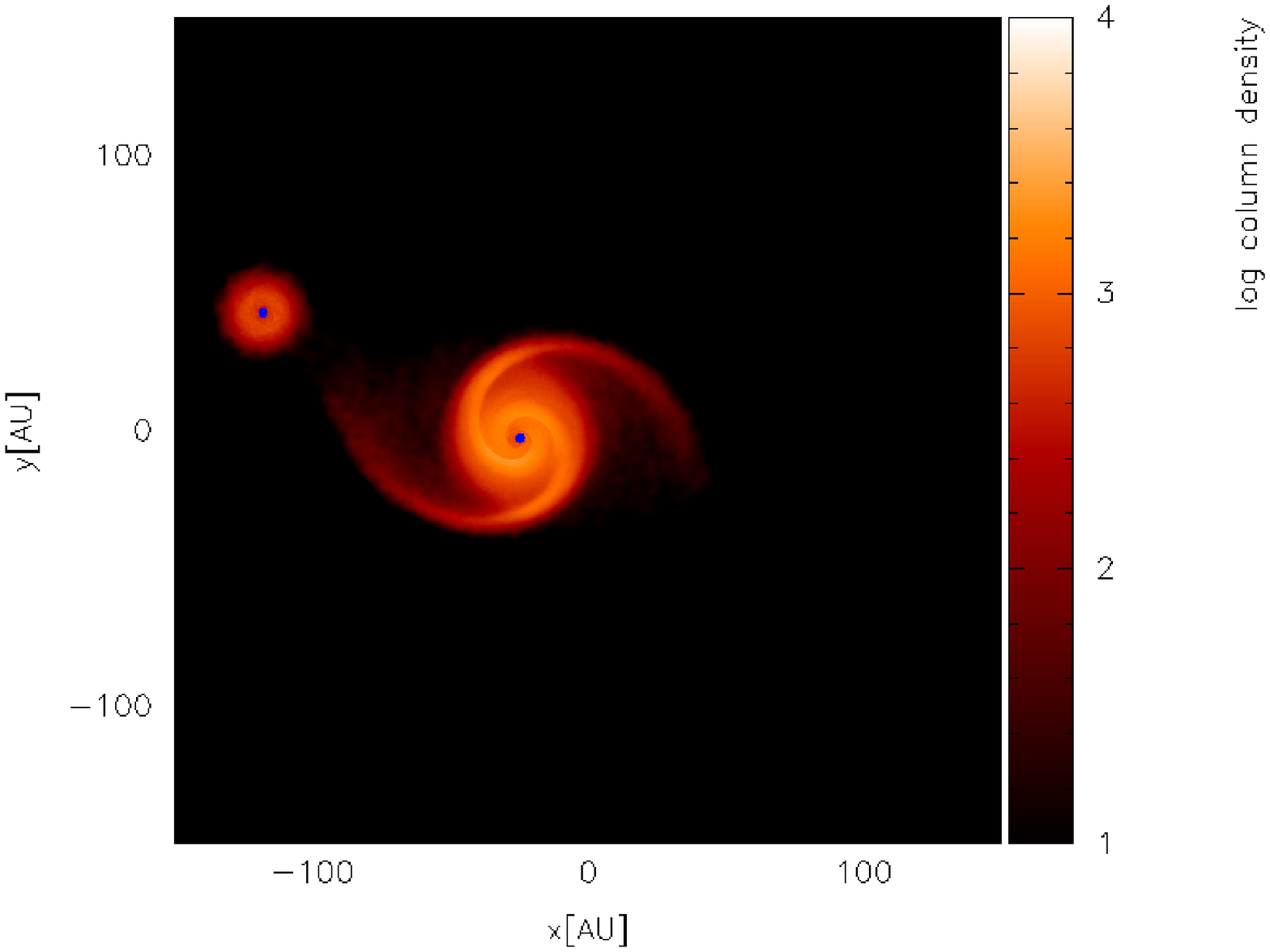} 
\end{array}$
\caption{Surface density plots of the $M_{\rm cl} = 1M_{\rm \odot}$
  simulations (Simulations 1, 2, 3, 4, 5 and 6).  The snapshots were
  taken two free fall times after the collapse was initiated (for Simulations 1 and 2, top), or at the point of disc fragmentation (for Simulations 3, 4, 5 and 6, middle left and right, and  bottom left and bottom right respectively).}\label{fig:Mcl1_surf}
\end{center}
\end{figure*}

Figure \ref{fig:Mcl1_surf} shows surface density plots for Simulations
1, 2, 3, 4, 5 and 6, after at least $2 t_{ff}$ of evolution, or
during fragmentation in the case of Simulations 3, 4, 5 and 6.  In
Simulations 3, 4 and 5, fragmentation occurs around $t=1.5t_{ff}$, in
Simulation 6 it occurs around $t=1t_{ff}$.  It is clear that increasing the initial cloud angular momentum increases the overall disc extent.  There is also a general tendency for fragments to form closer to the central object as the angular momentum is increased.  In all cases, it is apparent that the spiral structure is dominated by $m=2$ spiral modes - we shall return to the nature of their angular momentum transport in the following sections.

\begin{figure*}
\begin{center}$
\begin{array}{cc}
\includegraphics[scale = 0.4]{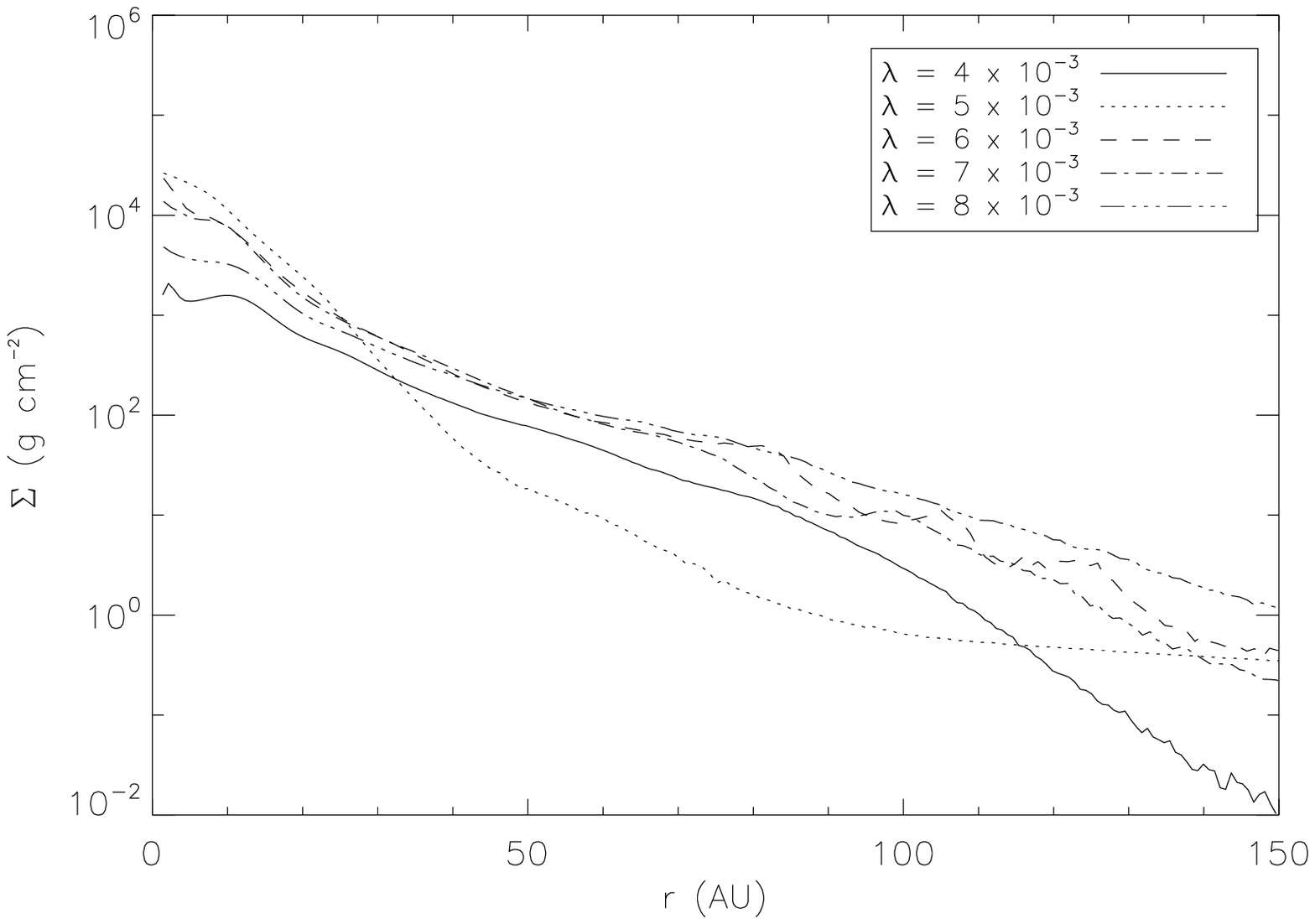} &
\includegraphics[scale = 0.4]{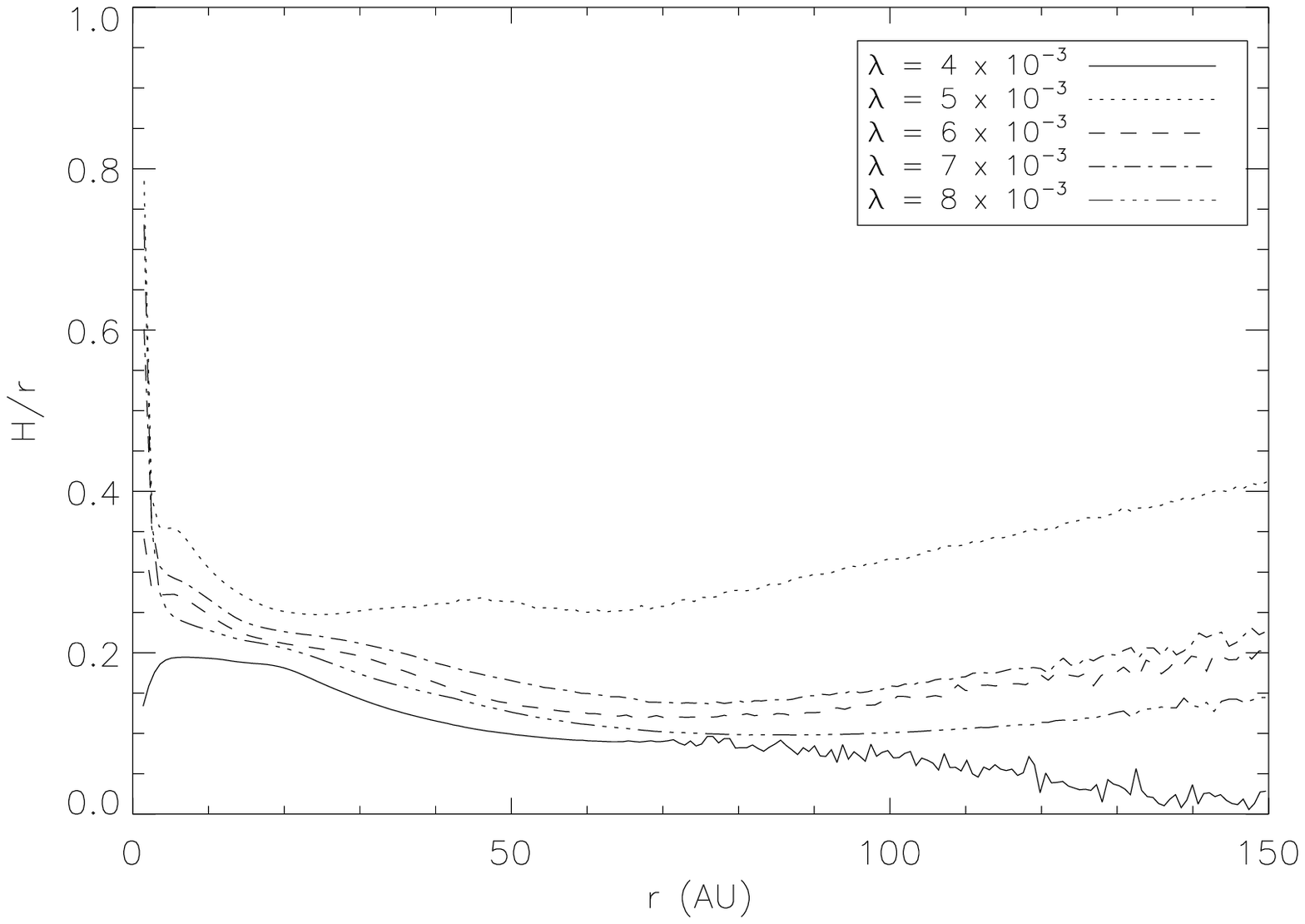} \\
\includegraphics[scale = 0.4]{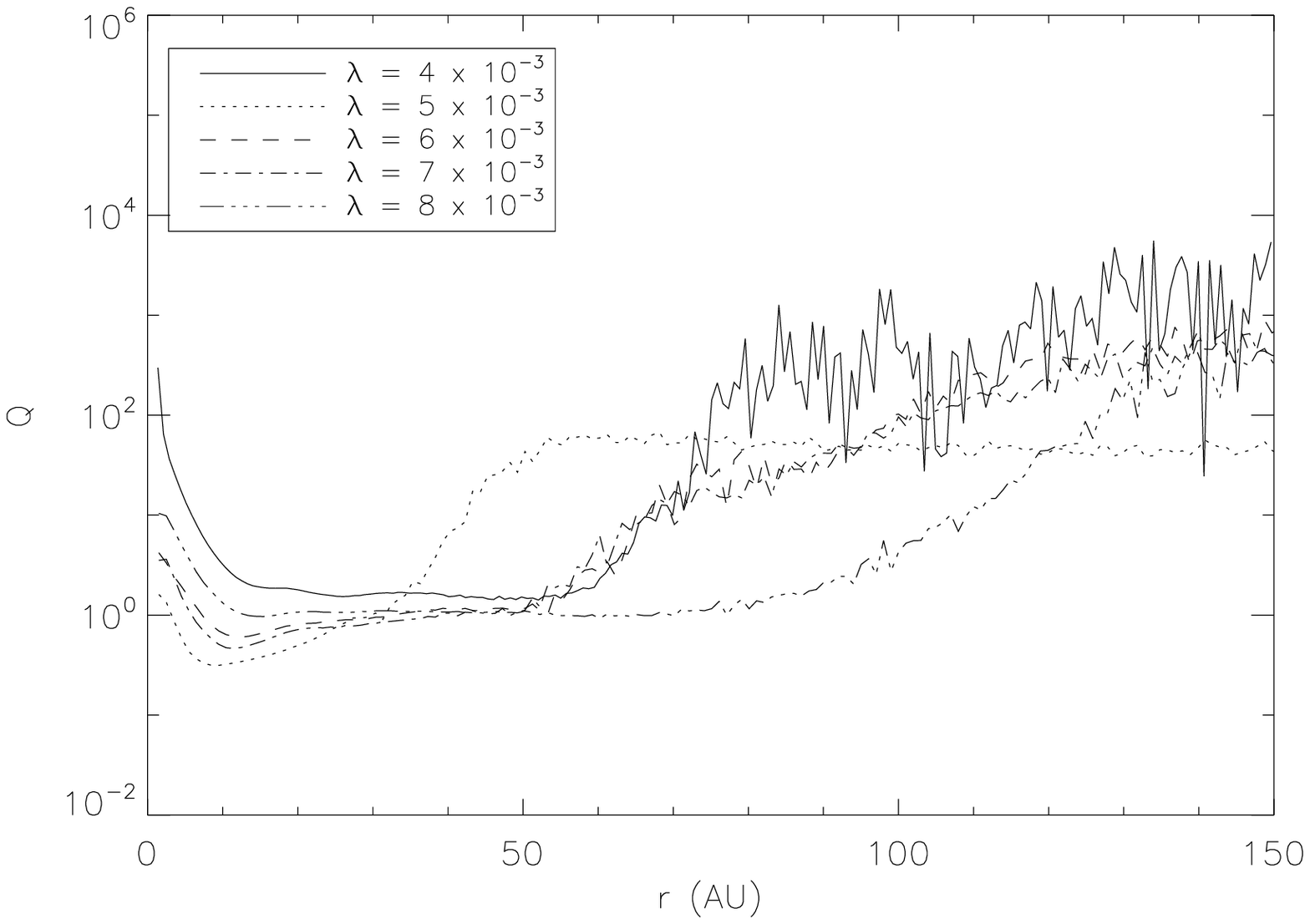} &
\includegraphics[scale = 0.4]{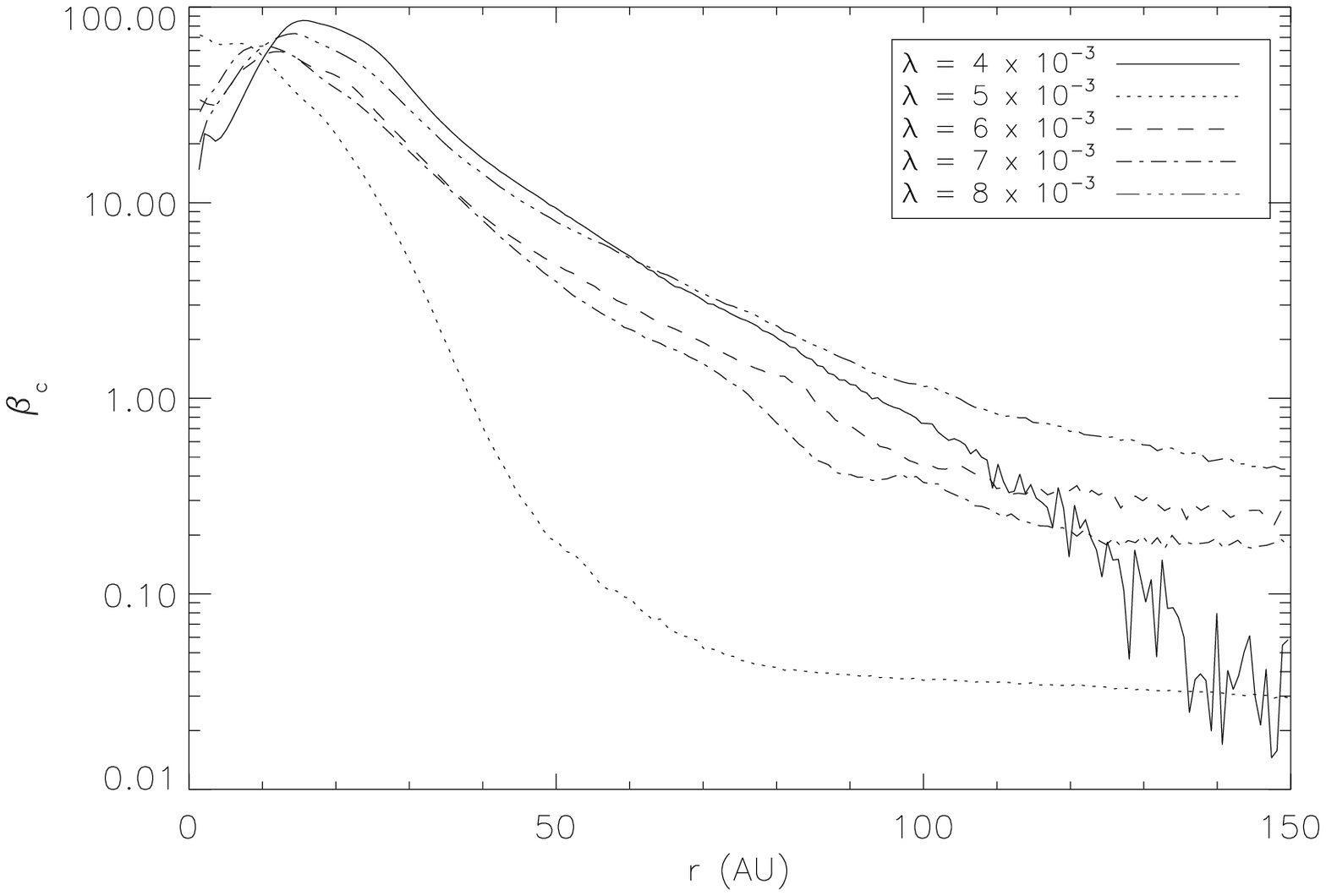} \\
\end{array}$
\caption{Azimuthally averaged radial profiles from the $M_{\rm cl} = 1
  M_{\rm \odot}$ simulations (Simulation 1 (solid line), Simulation 2
  (dotted lines), Simulation 3 (dashed lines), Simulation 4
  (dot-dashed lines) and Simulation 5 (triple-dot-dashed lines)).
  Simulation 6 is excluded as it forms a multiple system.  The
  profiles are time averaged from the formation of the disc to the
  point of fragmentation (in the case of Simulations 1 and 2, they are
  time averaged for approximately 7000 years, corresponding to
  approximately 19 ORPs).  The top left panel shows the surface
  density profile (calculated from the material contained within the
  first three scale heights only), the top right shows the aspect
  ratio, the bottom left shows the Toomre Q parameter, and the right
  hand panel shows $\beta_{c}$.}\label{fig:Mcl1}
\end{center}
\end{figure*}

Figure \ref{fig:Mcl1} shows azimuthally averaged radial profiles for
Simulations 1, 2, 3, 4 and 5.  The impact of increasing angular momentum
is evident in the surface density profiles (the top left panel of
Figure \ref{fig:Mcl1}).  

There appears to be a clear difference in trend between
non-fragmenting discs (Simulations 1 and 2) and fragmenting discs
(Simulations 3,4 and 5).  In the non-fragmenting case, the surface
density profile is visibly steeper (with a corresponding higher aspect
ratio, seen in the top right panel).  As these plots are
time-averaged, we can use each snapshot to derive the standard
deviations of each curve for all radii, and perform $\chi^2$ fitting.
All the curves are fitted to a power-law with exponential decay, of
the form \citep{Williams2011}:

\begin{equation} \Sigma(r) = \frac{(2-p)M_{\rm d}}{2 \pi R^2_c} \left(\frac{r}{R_c}\right)^{-p} \end{equation}

\noindent (where $p$ and $R_c$ are free parameters).  This yields
$p\sim 1.99$ for Simulation 1 and 2, and $p\sim 1$ for the other simulations.   These results are reflected in the Toomre $Q$ profiles in the bottom left panel, where Simulation 2 displays a very narrow range of gravitationally unstable radii in comparison to the other discs.  Indeed, despite their apparent similarities in the rest of Figure \ref{fig:Mcl1}, Simulations 3, 4 and 5 become distinct in their $Q$-profiles (although 3 and 4 remain similar).  Simulation 5 displays the largest region of instability, extending over a range spanning approximately $80$ au.

This set of results is an excellent example of disc fragmentation
requiring at least two criteria be met.  The bottom right panel shows
the dimensionless cooling time parameter $\beta_c$, which again
displays differing trends between fragmenting and non-fragmenting
discs.  Values of $\beta_c$ of order unity present amenable conditions
for fragmentation, and all four discs satisfy this condition in their
outer edges.  Simulation 2 satisfies this at the lowest
radii, at around 40 au.  However, we must also note that at 40 au,
$Q\sim 3$ and is increasing with radius.  Therefore, while this region
can maintain very cool gas, it is not gravitationally unstable, and we
therefore do not expect it to fragment - this expectation is
satisfied, despite running the simulation for 30,000 years.  As a test, we ran the disc for an extra 30,000 years, and no fragmentation was seen during this subsequent period.  

\begin{figure*}
\begin{center}$
\begin{array}{ccc}
\includegraphics[scale=0.4]{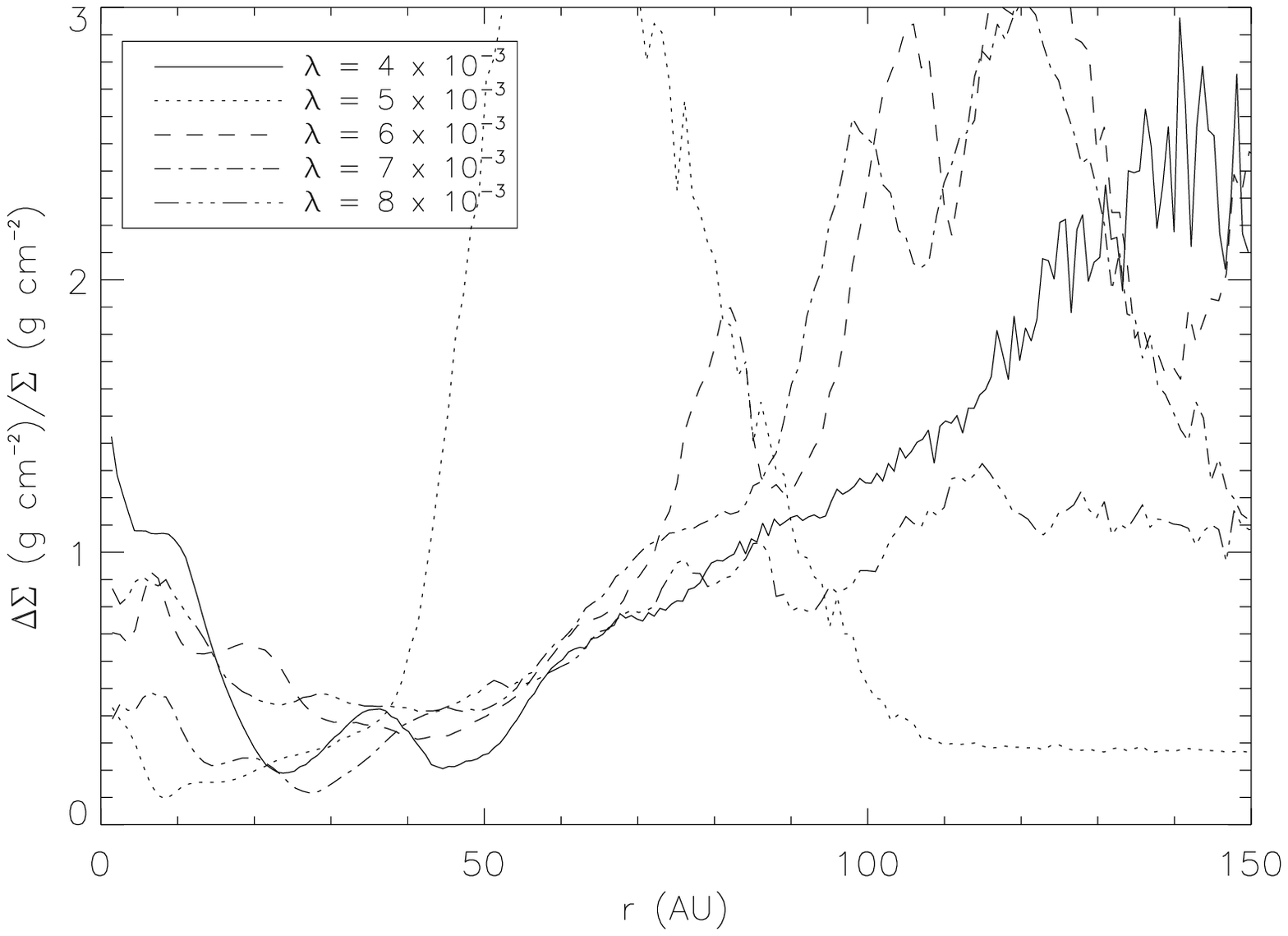} &
\includegraphics[scale = 0.4]{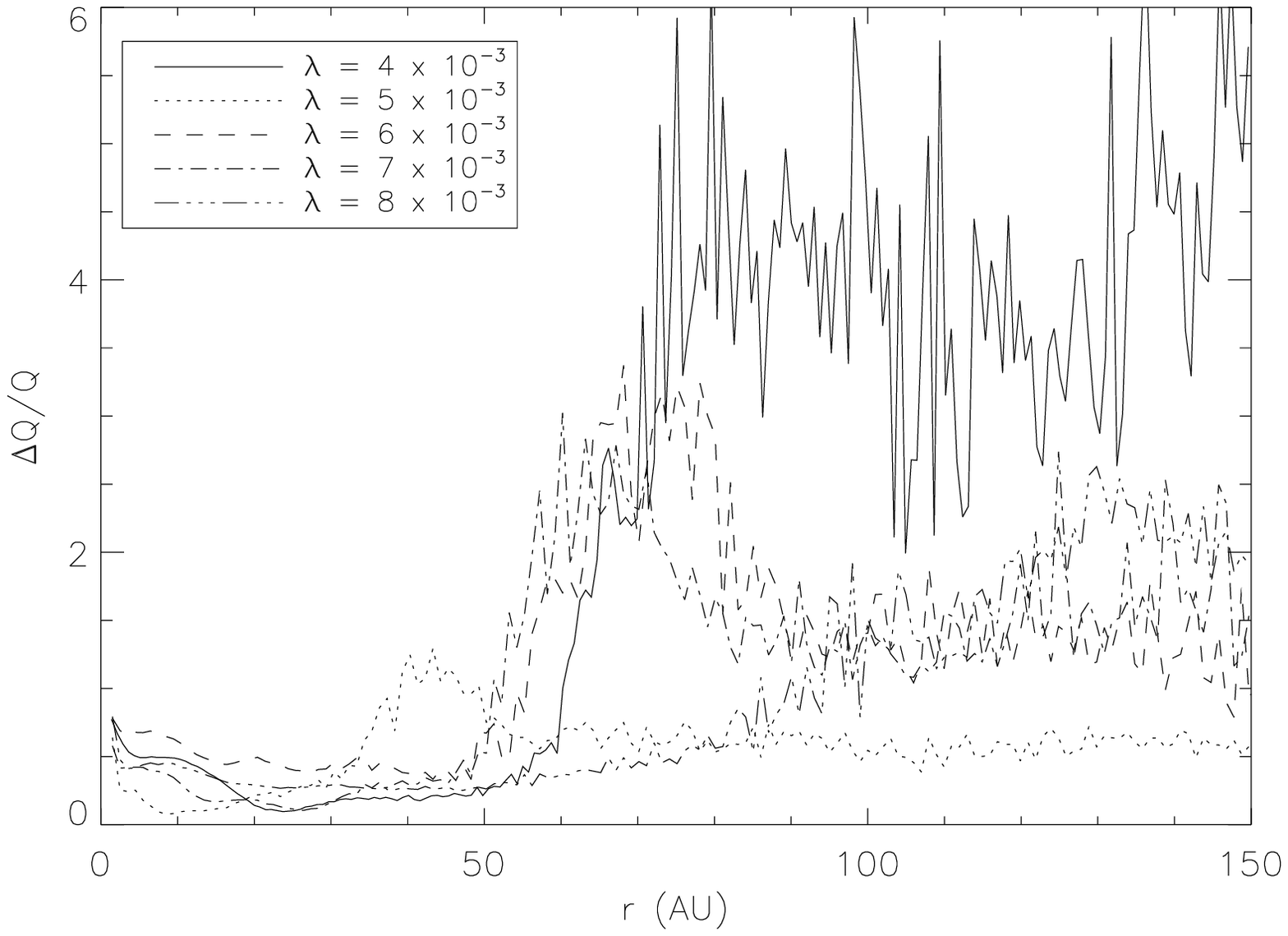} \\
\end{array}$
\caption{Variation in the mean surface density (left) and Toomre parameter Profile (right), for the
  $M_{\rm cl} = 1.0 M_{\rm \odot}$ simulations (1, 2, 3, 4 and 5).}\label{fig:delta_Mcl1}
\end{center}
\end{figure*}

The observant reader will note that fragments are able to form in
Simulations 3 and 4 at large radii, despite high values of $Q$.
Figure \ref{fig:delta_Mcl1} goes some way to explaining this
phenomenon.  These discs are highly variable (Simulation 2 being the
most steady-state of the four).  While temperature fluctuations in
these discs are relatively low, the fluctuations in $Q$ are quite
high, as has been observed previously in discs with infall
\citep{Harsono2011}.  Simulation 1 maintains fluctuations of order
50\% in the outer regions, ensuring $Q>>2$.   The other simulations
have extremely high $Q$ fluctuations (nearly 100\%), and all possess very low values of $\beta_c$, of order unity or less.  It is this transient behaviour that
allows $Q<2$ for a sufficient period of time (say one orbital period
at a given radius) which allows the disc to fragment.  Simulation 1
also has extremely variable $Q$, but this translates into relatively
weak surface density perturbations in regions of low $\Sigma$ (as can be seen in the left panel of Figure
\ref{fig:delta_Mcl1}), and therefore does not fragment.

In effect, for Simulations 3 and 4 to fragment, the square of the
epicyclic frequency must become negative, a phenomenon that typically
defines the outer edge of accretion discs, where perturbations from
circular flow will be unstable.  This tends to agree with the
consensus view that fragmentation in the inner regions of discs is
difficult to achieve \citep{Rafikov_05, Matzner_Levin_05,
  Whit_Stam_06,Mejia_3,Stamatellos2008, intro_hybrid, Clarke_09}.  On
the other hand, Simulation 5 shows less variation than the others, and
satisfies the canonical fragmentation criteria (see next section) at around 100 au, where fragments are subsequently seen.  The initial fragment provides an extra torque to the spiral wave that spawned it, which results in consecutive fragmentation along the wave.  For example, the condensation in the lower right quadrant of the lower right panel of Figure \ref{fig:Mcl1_surf} subsequently fragments into several objects.  In all the simulations conducted, if multiple fragmentation occurs, it seems to occur as a result of an initial fragment's torques on the disc, with time delays between the first object and the second object of up to an orbital period.

\subsubsection{Angular Momentum Transport}

\begin{figure}
\begin{center}
\includegraphics[scale = 0.5]{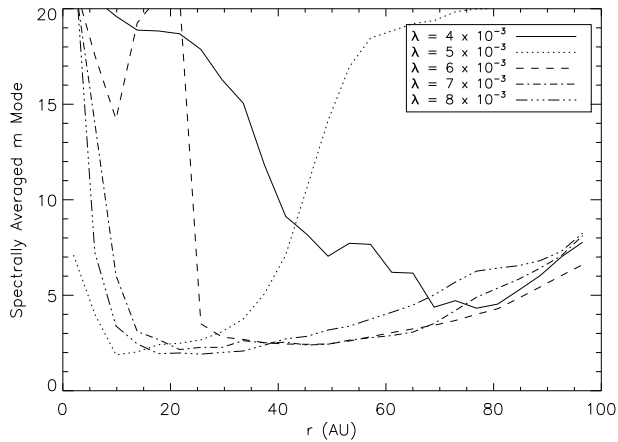}
\caption{Spectrally averaged $m$ modes as a function of radius for the
  $M_{\rm cl} = 1.0 M_{\rm \odot}$ simulations (1, 2, 3, 4 and 5).
  Note that inside 10-20 au the discs are dominated by numerical
  viscosity, meaning that data in this region should be ignored.  With the exception of Simulation 1, all discs exhibit
  non-local angular momentum transport.  Consequently, the $m=2$ mode
  dominates consistently at large radii for Simulations 2-5.}\label{fig:m_Mcl1}
\end{center}
\end{figure}

\noindent We have seen that the spiral structure appears to be
dominated by $m=2$ modes, and that the discs are highly variable.
These are both symptoms of non-local angular momentum transport
\citep{Forgan2011}, and have been seen in previous simulations with
envelope accretion \citep{Kratter2008,Harsono2011}.  We carried out a
Fourier decomposition to study the amplitude of the $m$-modes.  Figure
\ref{fig:m_Mcl1} shows spectrally averaged $m$-modes as a function of
disc position for Simulations 1-5. These confirm our visual inspection
of the discs' spiral structure - the dominant $m$-mode in the unstable
region of all five discs is $m=2$, with the majority of the power
remaining in the lower modes at larger radii.  However, Simulation 1
shows moderate power across a wide range of mode numbers\footnote{Spectral
  averaging weights the amplitude of each mode by $m$, giving rise to
  seemingly high average $m$ modes as shown here - it is likely that
  $m \approx 4$ modes dominate in the case of Simulation 1.}, suggesting that it
may not be subject to non-local transport.  This is confirmed by calculating the
non-local transport fraction $\xi$ \citep{Cossins2008,Forgan2011}.  All
simulations except Simulation 1 exhibit values above 0.5, indicating
non-local transport is significant.  By contrast, Simulation 1
maintains values less than 0.2, indicating the transport is indeed local.

\begin{figure*}
\begin{center}$
\begin{array}{cc}
\includegraphics[scale = 0.4]{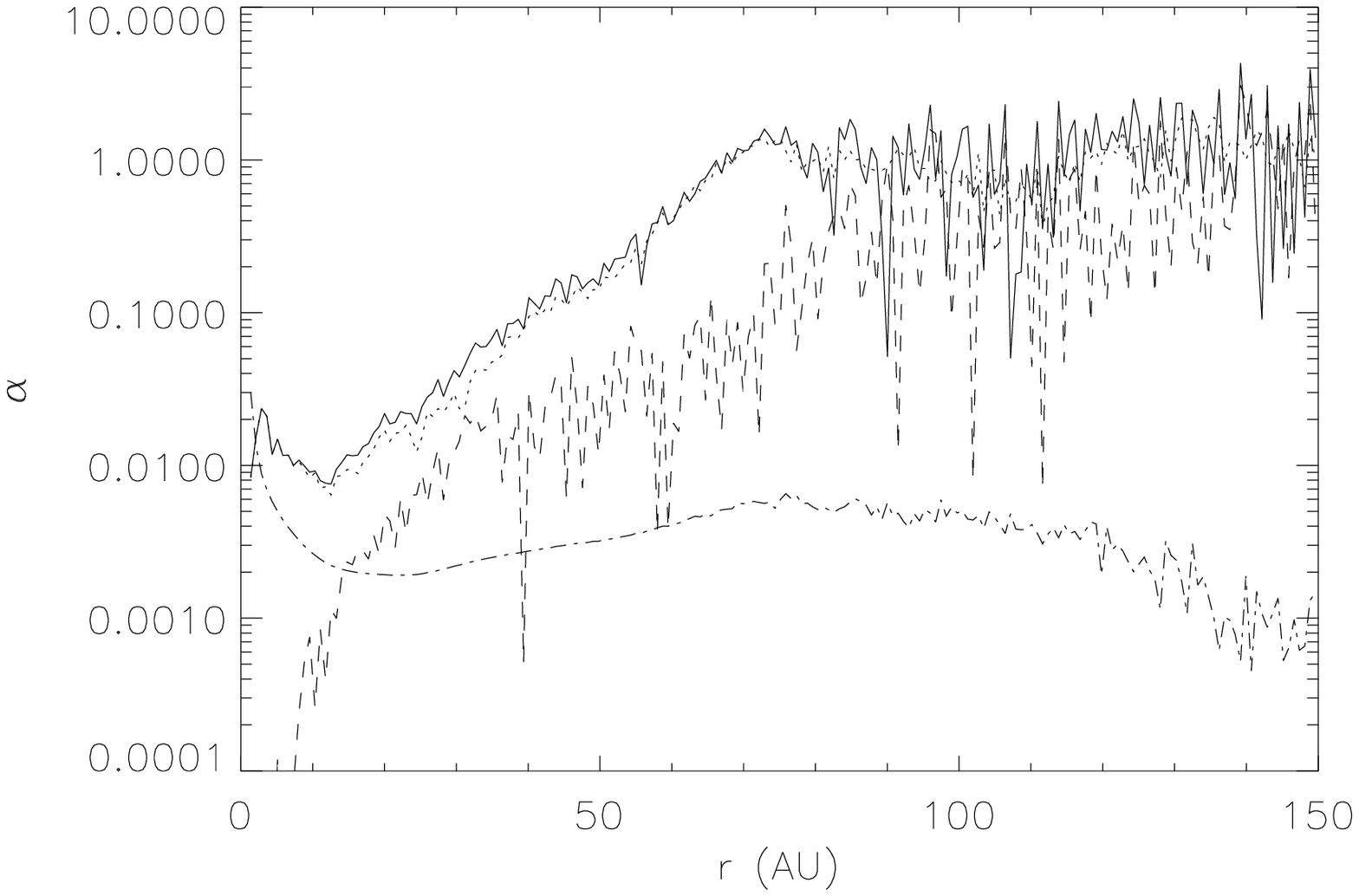}\\
\includegraphics[scale = 0.4]{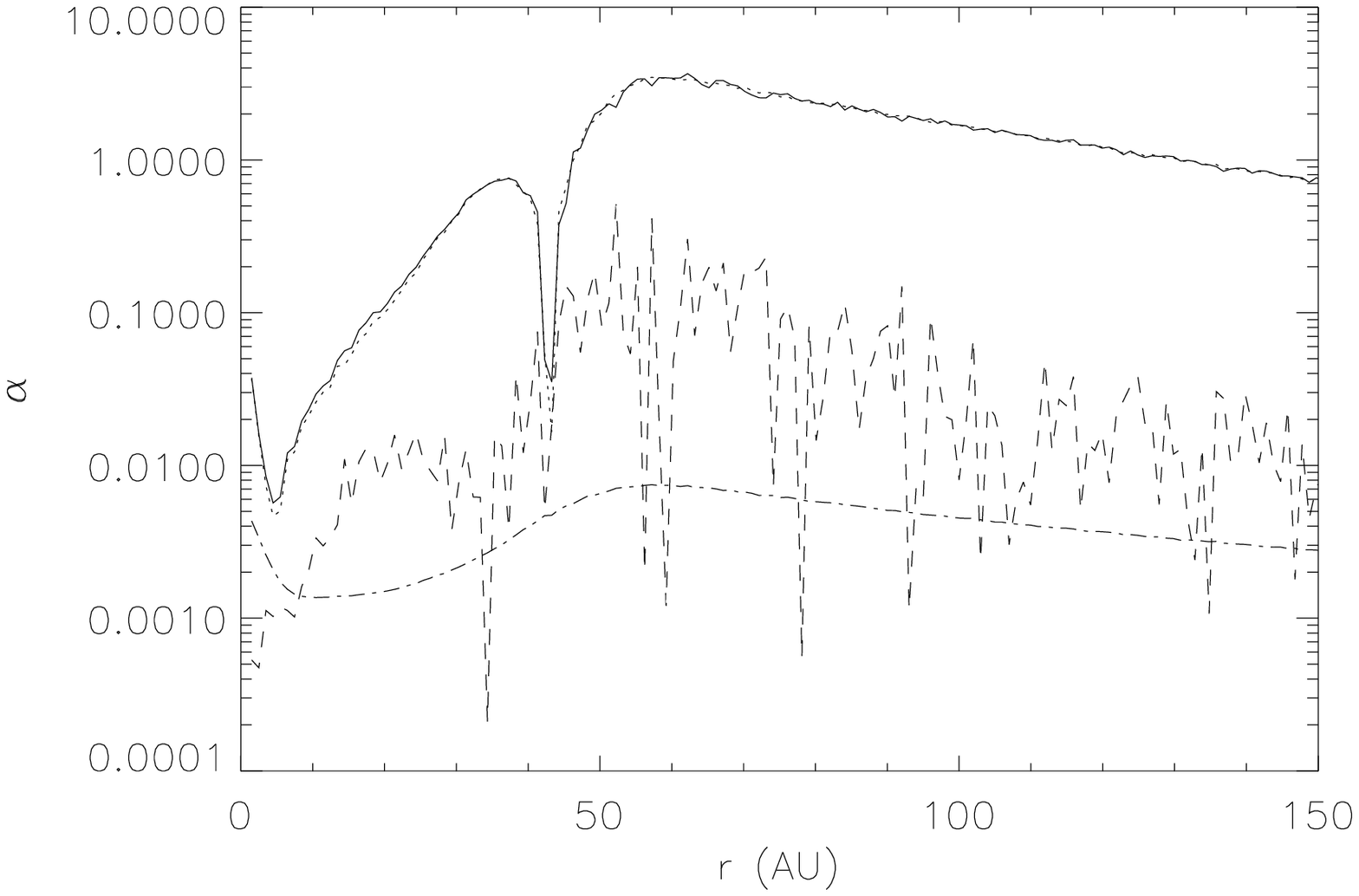} &
\includegraphics[scale = 0.4]{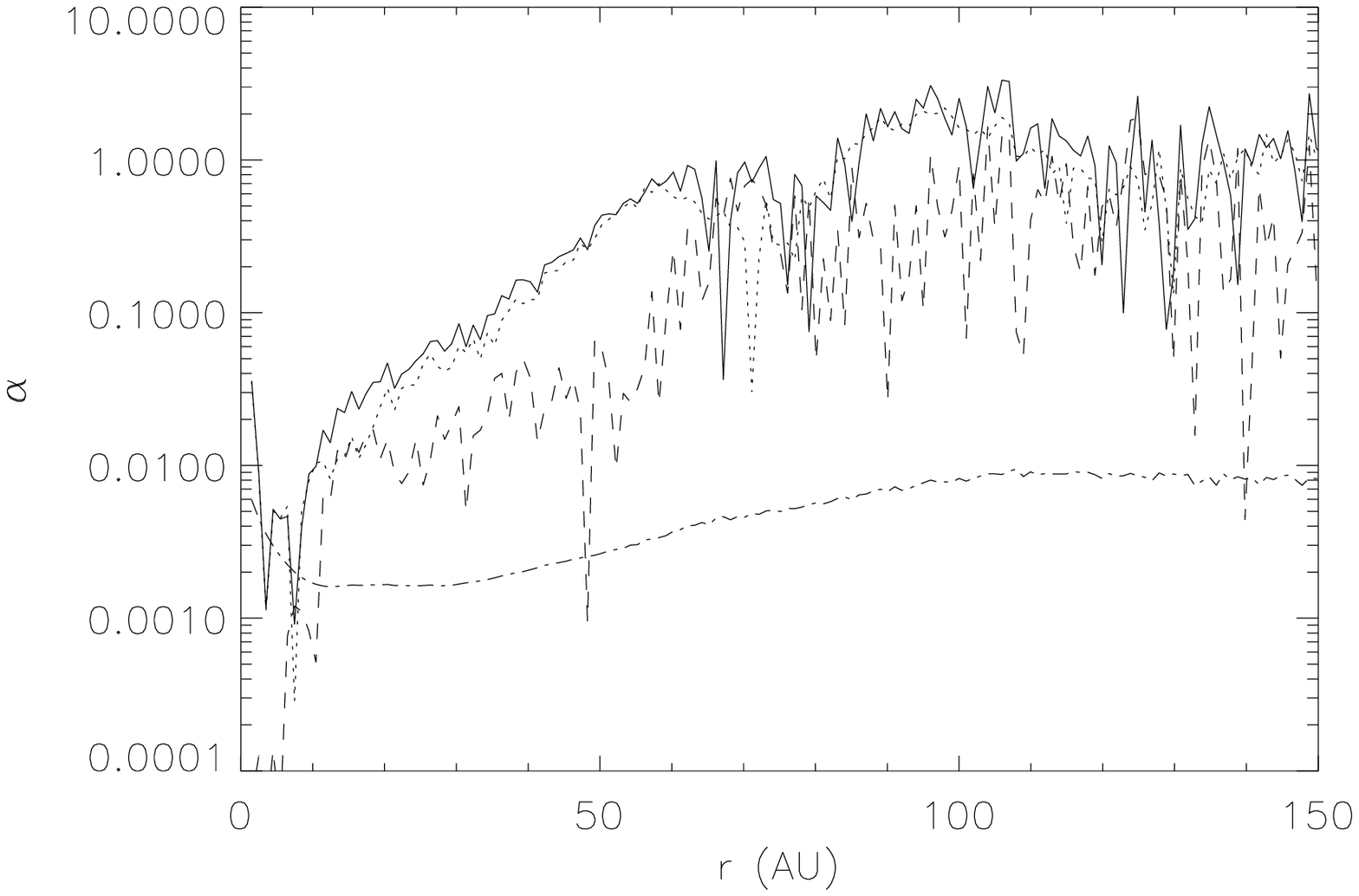} \\
\includegraphics[scale = 0.4]{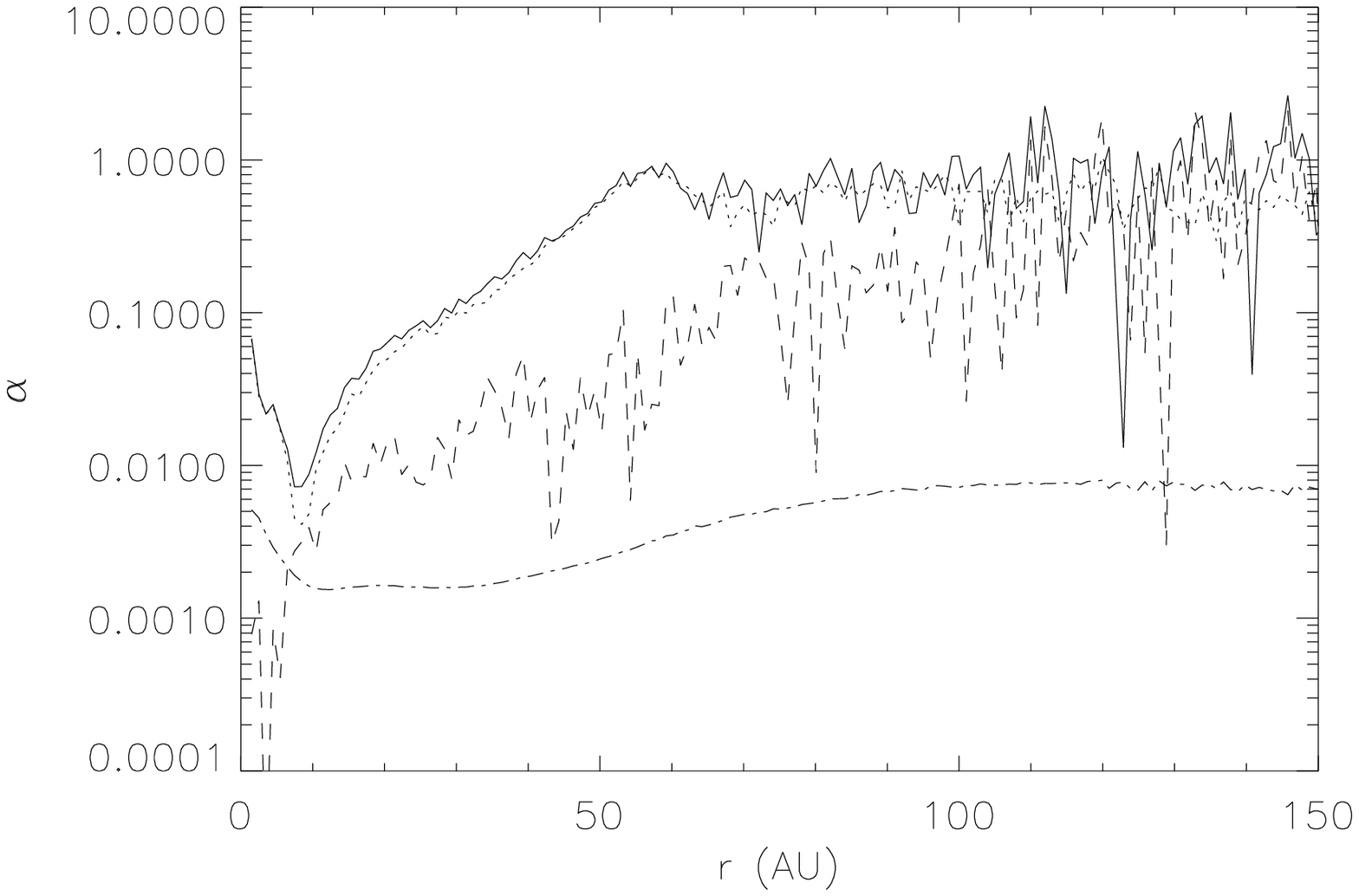} &
\includegraphics[scale = 0.4]{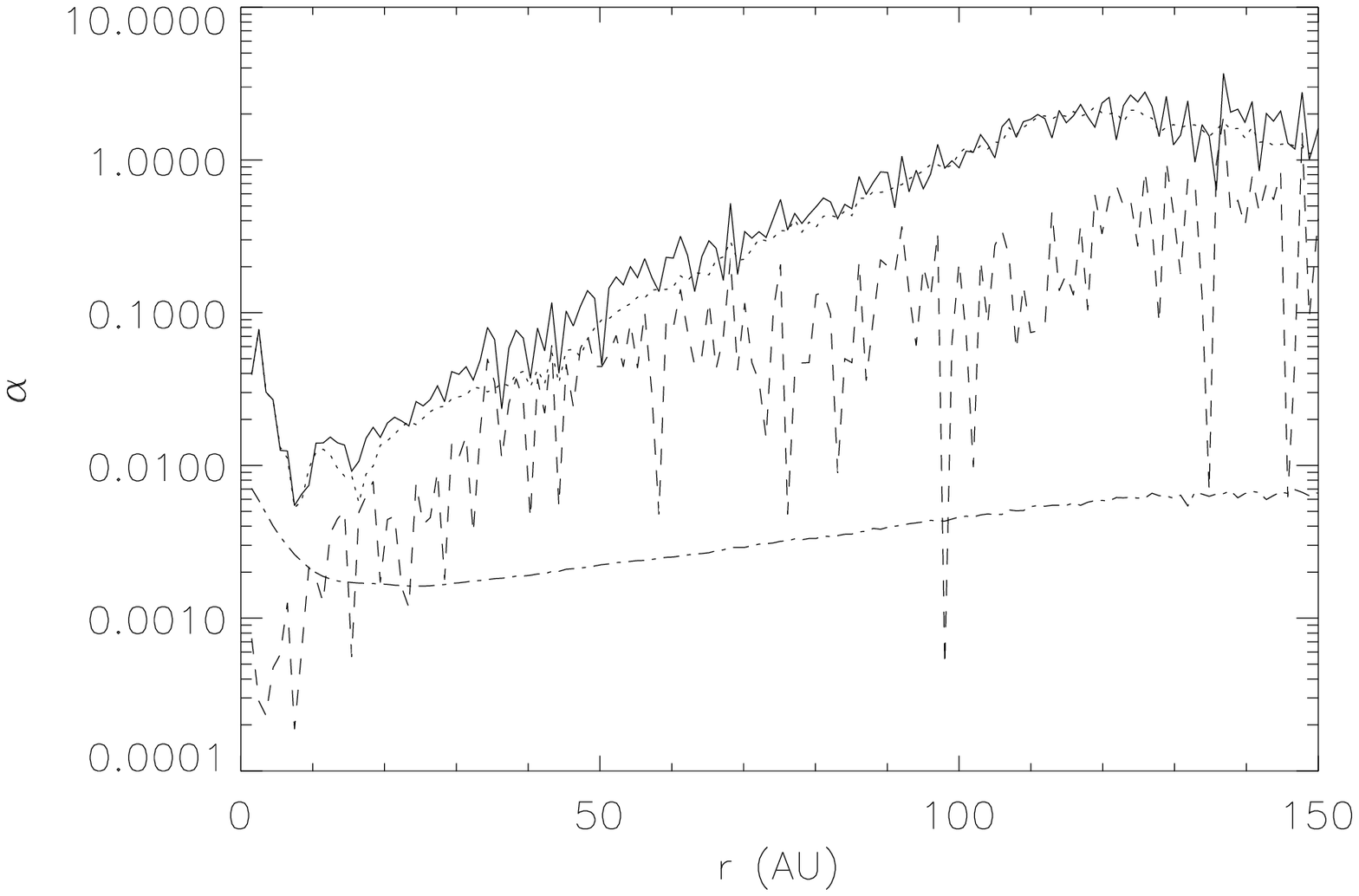} \\
\end{array}$
\caption{Azimuthally averaged $\alpha$-parameter for the $M_{\rm cl} = 1 M_{\rm \odot}$ simulations (Simulation 1 (top left panel), Simulation 2 (top right panel), Simulation 3 (bottom left panel) and Simulation 4 (bottom right panel)).  Simulation 5 is excluded as it forms a multiple system.  The profiles are time averaged from the formation of the disc to the point of fragmentation (in the case of Simulation 1, it is time averaged for 7000 years, corresponding to approximately 19 ORPs).  The solid line shows the total $\alpha$, the dashed lines show $\alpha_{grav}$, the dotted lines show $\alpha_{Reyn}$, and the dot-dashed lines show the contribution to the stress from artificial viscosity, $\alpha_{art}$.  Note that this numerical $\alpha$ dominates within around 10 AU, so we must disregard data from within this region.}\label{fig:alpha_Mcl1}
\end{center}
\end{figure*}

Figure \ref{fig:alpha_Mcl1} shows azimuthally averaged profiles of the $\alpha$-parameter, calculated for various components of the disc stress.  The solid lines indicate the total disc stress, the dashed lines show the contribution from gravitational stresses $\alpha_{\rm grav}$, and the dotted lines show the Reynolds stress $\alpha_{\rm Reyn}$ (expressions for these functions can be found in \citealt{Forgan2011}).  We also plot the stress induced by the artificial viscosity, $\alpha_{\rm art}$ to confirm that it is small in comparison to the other disc stresses.  

In contrast to isolated disc simulations, we can see that the Reynolds stresses are typically large, of order $\alpha \sim 1$.  This is in keeping with values obtained by \citet{Kratter2008} and \citet{Harsono2011} in similar disc configurations.

Again, there is a stark contrast between non-fragmenting and
fragmenting discs.  We can see that the standard $\alpha_{\rm
  tot}=\alpha_{\rm grav} + \alpha_{\rm Reyn} >\alpha_{\rm crit} = 0.06$ criterion (derived for isolated discs) is invalid.  The large Reynolds stresses (induced by velocity shear from infalling envelope material contacting the disc's rotating flow) prevent the gravitational stress from dominating unless it can reach values closer to unity (in some cases up to 5 times larger than $\alpha_{\rm crit}$).  We should therefore be careful about using this criterion in young systems that are still surrounded by a substantial envelope - indeed, we should develop new criteria that take this accretion into account \citep{Forgan2011a}.

From this data, we suggest that fragmenting discs require the local
gravitational stress to be of the order (or larger than) the local
Reynolds stress.  We can see that Simulation 2 satisfies this
condition weakly at 40 au, again where $Q$ is too large for the disc
to be sufficiently unstable. At larger radii, the gravitational stress
decreases with distance, further ensuring no fragmentation can occur.
Simulation 1 appears to satisfy this at
100 au, but $\alpha_{\rm grav}$ varies quite strongly, only dominating
at much larger radii, where the surface density becomes negligible.
The differences between Simulations 1 and 2 are due to the different
modes of angular momentum transport at play (local and non-local
respectively).

Simulations 3 and 4 show gravitational stresses comparable to local
Reynolds stresses at around 150 and 110 au respectively - these
correspond well to the initial formation radius of the
fragments. Simulation 5 appears to be qualitatively different, in that
it fragments in a regime where the gravitational stress is still small
compared to the Reynolds stress, but $\alpha_{\rm grav}>\alpha_{\rm
  crit}$.  The disc exhibits more power in higher $m$-modes than the
others, and the total $\alpha$-profile differs also.  Where
Simulations 3 and 4 show relatively flat $\alpha_{\rm tot}$ outside 50
au, Simulation 5 shows a steadily increasing curve.  Naively comparing with
\cite{Forgan2011}'s study of isolated discs, this would suggest that
Simulation 5 shows more evidence of local angular momentum transport
than Simulations 3 and 4.  However, as has been said, Simulations 3-5
all possess large non-local transport fractions.

Simulations 3-5 produce disc fragments at a minimum distance of
100 au from the central star, where the local (time-averaged)
$\alpha_{\rm grav}$ at the fragmentation radius ranges from 0.06 to
0.7.  The subsequent evolution of the fragments in the non-linear
regime vary between each simulation - for example, Simulations 3 and 5 produce several subsequent fragments each, as a result of torques induced by the presence of the initial fragment, which itself undergoes rapid inward migration.  Simulation 4 produces a single fragment, which grows to form a second pointmass, where the mass ratio $M_2/M_1 \sim 0.15$.  

\section{Discussion \& Conclusion}\label{sec:Discussion}

\noindent From a set of numerical experiments with simplified initial
conditions, we have shown that the initial cloud angular momentum tends to
be the critical factor in subsequent disc fragmentation (compared to
the initial thermal energy of the cloud).  We confirm the results of
\citet{Rice2010}, in that fragmentation does not occur if $\lambda<
10^{-3}$.  While they also state that fragmentation occurs if
$\lambda>10^{-3}$, their parameter study does not resolve the
fragmentation boundary to the same level that this paper does.  In
this sense, we can say that their conclusions are broadly correct.
Agreement can also be found on the fragmentation radius, which we also
see increasing with initial cloud mass, and the steep surface density
profiles obtained for non-fragmenting discs, placing a large amount of
material within 20 au (although on this second statement we must
caution that numerical effects dominate within 10 au).  This agreement
is somewhat surprising, in that they assume that the angular momentum
transport due to self-gravity is local, where it has been established
in this paper and in others that it is likely to be dominated by
global spiral modes (cf \citealt{Harsono2011}).

All fragments form at distances above 70 au, typically consistent with previous studies of both isolated and accreting discs \citep{Rafikov_05, Matzner_Levin_05, Whit_Stam_06,Mejia_3,Stamatellos2008, intro_hybrid, Clarke_09, Rice2010}, which indicate that, despite maintaining low values of $Q$, the inner regions do not cool efficiently or impart sufficient stress due to the gravitational instability to produce fragments.  This is true in spite of non-local angular momentum transport or accretion from the envelope.  It appears that maintaining surface density profiles which place large amounts of (cool) material at large enough radii is the key to fragmentation.  The disc-to-star mass ratios are typically of order 0.5 at fragmentation, although they can be lower (cf \citealt{Stamatellos2011}). The masses of fragments formed are typically range from 1 to 15 Jupiter masses: these are roughly consistent with the local Jeans mass in the spiral waves from which they form \citep{Forgan2011a}, except in some cases where the Jeans length is sufficiently long to allow the break up of fragments into multiple objects. Future work will focus on the survival of such fragments as they condense out of the disc, to investigate the role of internal pressure support \citep{Kratter2011} and the importance of fragment-fragment collisions \citep{Shlosman1989}.

Marginally unstable, self-gravitating discs which do not fragment can
remain massive for long periods of time.  In one case, the disc was
shown to remain at masses roughly half that of the parent star for
around 50,000 years after formation, showing that the self-gravitating
phase could be more long-lasting than previously thought, much as the
Class 0 and Class I timescales have also been shown to be much longer
than expected \citep{Evans2011}.

The inherent variability seen in all simulations conducted
indicates that accretion onto the central star varies significantly.
Even in the more quiescent, non-fragmenting simulations, the value of
$Q$ varies by up to 50\% or more, ensuring episodes of
gravitational instability marked by high-amplitude, low-$m$ spiral
waves followed by episodes of relative stability with weak spiral
structures.  This variability occurs despite the omission of stellar
radiative feedback physics, unlike e.g. \citet{Stamatellos2011a}, who
demonstrate that accretion luminosity can enhance this episodic nature
of instability. This would suggest that self-gravity alone may account for the variability seen in Class I protostars \citep{Boley_and_Durisen_09,Vorobyov2009}.  It is quite possible that the one-parameter boundary we find disappears with the addition of stellar luminosity, as its strength will clearly depend on $M_{\rm cl}$.  

Indeed, this simple boundary is unlikely to apply in nature due to
other complicating factors, including non-uniform density profiles and
the addition of turbulence.  A density profile that decays with
radius, such as a supercritical Bonnor-Ebert sphere is a more physical
choice, ensuring more mass will be concentrated in the inner disc at
early times.  We suggest that these steeper profiles will require an
increased amount of initial angular momentum to produce fragmentation,
and that the boundary established for uniform clouds is a lower
limit\footnote{Strictly, an even lower limit could exist if the density increased with radius, but these conditions are even less likely to appear in nature than simple uniform conditions}.

The addition of turbulence tends to weaken any correlation with angular momentum \citep{Walch2010}.  Spherical symmetry is broken by construction, and discs accrete from the environment through filamentary structures, allowing significant specific angular momentum to be injected asymmetrically, exciting low-$m$ spiral modes as the disc attempts to redistribute itself.  In general, discs formed from turbulent simulations tend to grow more slowly, and have smaller radii for a given angular momentum - this again suggests that the criterion derived in this paper remains a lower limit.  

Considering that the threshold angular momentum required ($\lambda \gtrsim 5 \times 10^{-3}$) is small compared to observations, which indicate a median of $\lambda \sim 0.02$ \citep{Goodman1993} - this would suggest that disc fragmentation will be a somewhat common process, but occur on such short timescales that observations of such events will be rare \citep{Stamatellos2011}.  Exactly how this threshold might vary with surface density profile and the addition of turbulence is difficult to quantify, but merits further research.

\section*{Acknowledgments}

\noindent Density plots were produced using SPLASH \citep{SPLASH}.  All simulations were performed using high performance computing funded by the Scottish Universities Physics Alliance (SUPA).  DF and KR gratefully acknowledge support from STFC grant ST/H002380/1.

\bibliographystyle{mn2e} % (must include a bibliography style)
\bibliography{collapses}

\appendix

\label{lastpage}

\end{document}